\documentclass{emulateapj}
\usepackage{graphicx}
\usepackage{float}
\usepackage{times}

\begin{document}

\slugcomment{The Astrophysical Journal, 590:619-635, 2003 June 20}
\shorttitle{Dissipative Formation of an Elliptical Galaxy}
\shortauthors{Meza et al.}

\title{Simulations of Galaxy Formation in a $\Lambda$CDM Universe III:
The Dissipative Formation of an Elliptical Galaxy}

\author{Andr\'es Meza and Julio F. Navarro\altaffilmark{1}}
\affil{Department of Physics and Astronomy, University of Victoria,
Victoria, BC V8P 1A1, Canada}
\author{Matthias Steinmetz\altaffilmark{2}}
\affil{Steward Observatory, 933 North Cherry Avenue, Tucson, AZ 85721,
USA, and  Astrophysikalisches Institut Potsdam, An der Sternwarte 16,
D-14482 Potsdam, Germany}
\and
\author{Vincent R. Eke\altaffilmark{3}}
\affil{Physics Department, Durham University, South Road, Durham DH1
3LE, England}
 
\begin{abstract}
We examine the dynamical structure of an elliptical galaxy simulated in
the $\Lambda$CDM scenario. The morphology of the galaxy evolves
dramatically over time in response to the mode and timing of mass
accretion; smooth deposition of cooled gas leads to the formation of
centrifugally supported disks, whilst major mergers disperse stellar
disks into spheroids. In the case we consider here, these two modes of
accretion alternate successively until $z\sim 0.6$, when the galaxy
undergoes one last major ($1$:$3$) merger. The event triggers a
starburst that consumes much of the remaining gas into stars. Little gas
cools and accretes subsequently and, as a result, most stars at $z=0$
are rather old ($75\%$ are older than $9$ Gyr), and distributed in a
spheroidal component that resembles present-day elliptical galaxies.
Dynamically, the galaxy is well approximated by an E4 oblate rotator,
with rotational support increasing steadily from the center outwards. 
The apparent rotation support, as measured by $V_{\rm rot}/\sigma$,
correlates strongly with isophotal deviations from perfect ellipses.
Boxy isophotes are obtained when the galaxy is seen face-on and $V_{\rm
rot}/\sigma \ll 1$. On the other hand, disky isophotes are found for
inclinations which maximize $V_{\rm rot}/\sigma$. The line-of-sight
velocity distribution is nearly Gaussian at all radii, although small
but significant deviations are robustly measured. The sign of the
Gauss-Hermite skewness parameter $h_3$ is anti-correlated with the
apparent sense of rotation, in agreement with observed trends. Despite
its relatively recent assembly, the simulated galaxy has much higher
effective surface brightness than normal ellipticals of similar
luminosity, in a way reminiscent of the less common type of M32-like
``compact ellipticals''. This is likely a direct consequence of our star
formation and feedback algorithm, which allows for efficient
transformation of gas into stars in dense, early collapsing progenitors
rather than a definitive prediction for the structure of galaxies
assembled in this $\Lambda$CDM halo. Despite this limitation, our
simulation suggests that dark matter plays a minor role in the luminous
regions of compact ellipticals, whose dynamical mass-to-light ratios are
thus not dissimilar to those of normal ellipticals. This explains the
proximity of compact ellipticals to edge-on projections of the
Fundamental Plane, despite their far greater velocity dispersion at
given luminosity. Overall, our simulation shows that repeated episodes
of dissipational collapse, followed by merger events, lead to stellar
spheroids that are only mildly triaxial and of relatively simple
kinematic structure. This is in better agreement with observation than
earlier models based on dissipationless mergers of stellar disks, and a
positive step towards reconciling the observed structure of ellipticals
with a hierarchical assembly process where mergers play a substantial
role.
\end{abstract}


\altaffiltext{1}{Fellow of the Canadian Institute for Advanced Research
and of the Alfred P. Sloan Foundation}
\altaffiltext{2}{Packard Fellow and Sloan Fellow}
\altaffiltext{3}{Royal Society University Research Fellow}

\section{Introduction}
\label{sec:intro}

Elliptical galaxies constitute a rather uniform family of stellar
spheroids whose dynamical and photometric properties satisfy a number of
scaling relations and whose detailed internal structure appear to vary
weakly but systematically with galaxy mass. Perhaps the most important
of such scaling relations is the Fundamental Plane; a tight correlation
between the luminosity, size, and velocity dispersion that has been
extensively used as a sensitive distance indicator (Djorgovski \& Davis
1987; Dressler et al. 1987). The existence of the Fundamental Plane is
believed to reflect largely the virial theorem, as applied to systems
where the dark matter content, as well as perhaps age and metallicity,
vary monotonically as a function of luminosity.

Subtler dynamical and photometric indicators also vary with luminosity;
bright ellipticals are typically slowly-rotating triaxial structures
supported by anisotropic velocity dispersion tensors, whereas rotation
is thought to play a more substantial role in the dynamics of fainter
ellipticals (Davies et al. 1983; Bender \& Nieto 1990; Halliday et al.
2001). This dynamical distinction is reflected in isophotal deviations
from perfect ellipses: fast-rotating spheroids are usually ``disky'', in
contrast with the ``boxy'' isophotal shapes of slowly rotating systems
(Bender 1988).  Such deviations---although significant and robustly
measured---are usually small, and simple concentric ellipsoidal models
provide a rough but reasonable description of the structure of many
elliptical galaxies. Indeed, detailed dynamical analysis indicates that
most ellipticals are globally only mildly triaxial, as indicated, for
example, by the weak misalignment between photometric and dynamical
principal axes (de Zeeuw \& Franx 1991; Franx, Illingworth, \& de Zeeuw
1991).

These structural trends hold important clues to the formation and
evolution of ellipticals, and are not straightforward to account for in
traditional models of elliptical galaxy formation. In the ``monolithic
collapse'' model put forward by Partridge \& Peebles (1967), all stars
in an elliptical galaxy form coevally at very early times as a result of
the coherent, dissipationless collapse of a large mass of gas
transformed into stars in a short and intense bout of star formation.
This simple model accounts naturally for the old apparent ages of
spheroid stars, for their high densities, and for the weak evolution in
the spheroidal population properties with time, but is less successful
at explaining the detailed luminosity dependence of their dynamical
properties; the apparent scarcity of very large starbursts in the
high-redshift universe; and the origin of dynamical peculiarities
suggestive of recent accretion events (e.g., Franx \& Illingworth 1988;
Jedrzejewski \& Schechter 1988; Surma \& Bender 1995; de Zeeuw et al.
2002).

In its most extreme form, the competing ``hierarchical assembly'' model
posits that spheroidal galaxies form through the violent merger of
preexisting disk galaxies (Toomre 1977). In this scenario, the epoch of
assembly of ellipticals differs markedly from the epoch of formation of
their constituent stars, and the high density of ellipticals is ascribed
to the effects of dissipation during the formation of the progenitor
disks. This model accounts ``naturally'' for the scarcity of very bright
elliptical progenitors at high-redshift; for the rapid evolution of the
galaxy population with lookback time; and for the presence of dynamical
peculiarities. It is, on the other hand, less successful at explaining
the apparent old ages of stars in ellipticals and the uniformity in
their dynamical properties. Indeed, it has been plausibly argued that
the remnants of major mergers between stellar disks cannot reproduce the
steep central surface brightness profiles of ellipticals, and that
dissipationless merger remnants are too triaxial, anisotropic, and
kinematically peculiar to be consistent with normal ellipticals (Barnes
1992; Hernquist 1992, 1993).

One possible solution to these problems is to argue that the
hierarchical mergers that lead to the formation of ellipticals are
gas-rich, and that substantial dissipation takes place during the
collision, where renewed star formation of gas driven to the center can
steepen the luminosity profiles and even out misalignments in the
dynamical structure. This possibility has been explored by Mihos \&
Hernquist (1994; and references therein) who find that major mergers may
actually be too efficient at funneling gas towards the center, leading
to extremely dense stellar cores in disagreement with observation. 

Taken together, this body of work suggests that it might indeed be
possible to reconcile the observed structural properties of ellipticals
with the hierarchical formation scenario, but only provided that star
formation somehow regulates effectively the flow of accreting gas within
galaxies, especially during mergers. Thus, accounting self-consistently
for the frequency of merging---which is determined largely by the
cosmological context of formation---is crucial for the success of the
modeling and has yet to be taken into account properly in simulation
work.

We present here the result of our first attempt at simulating the
formation of an elliptical galaxy {\it ab initio} within the popular
$\Lambda$CDM cosmogony. The simulation evolves self-consistently a
region destined to aggregate into a single dark matter halo and includes
the gravitational and hydrodynamical effects of dark matter and baryons,
photo-heating by a UV background, Compton and radiative cooling, as well
as star formation and feedback processes. The main evolutionary features
of this simulation have been presented by Steinmetz \& Navarro (2002). 
This is one of the new generation of simulations able to resolve the
internal dynamics of galaxies whilst accounting properly for the
cosmological context of the formation process (see also Abadi et al.
2003a,b; Governato et al. 2002; Sommer-Larsen et al. 2002a,b). In this
paper, we focus on the $z=0$ structure and kinematics of the stellar
component of the simulation reported by Steinmetz \& Navarro (2002).

Major aims of this paper are to validate the analysis techniques that we
intend to apply to a statistically meaningful set of simulations (under
preparation) of comparable numerical resolution, as well as to identify
potential disagreements between simulation and observation that may
prompt a radical revision of our simulation algorithm or of the adopted
cosmological model. The plan of this paper is as follows. 
Section~\ref{sec:nummeth} presents a brief description of our numerical
techniques, whereas \S~\ref{sec:results} presents the main results of
the simulation. We summarize our findings in \S~\ref{sec:summary}.

\section{Numerical Methods}
\label{sec:nummeth}

\subsection{The Code}

\begin{figure*}[tb]
\centering\includegraphics[height=0.73\textheight,clip]{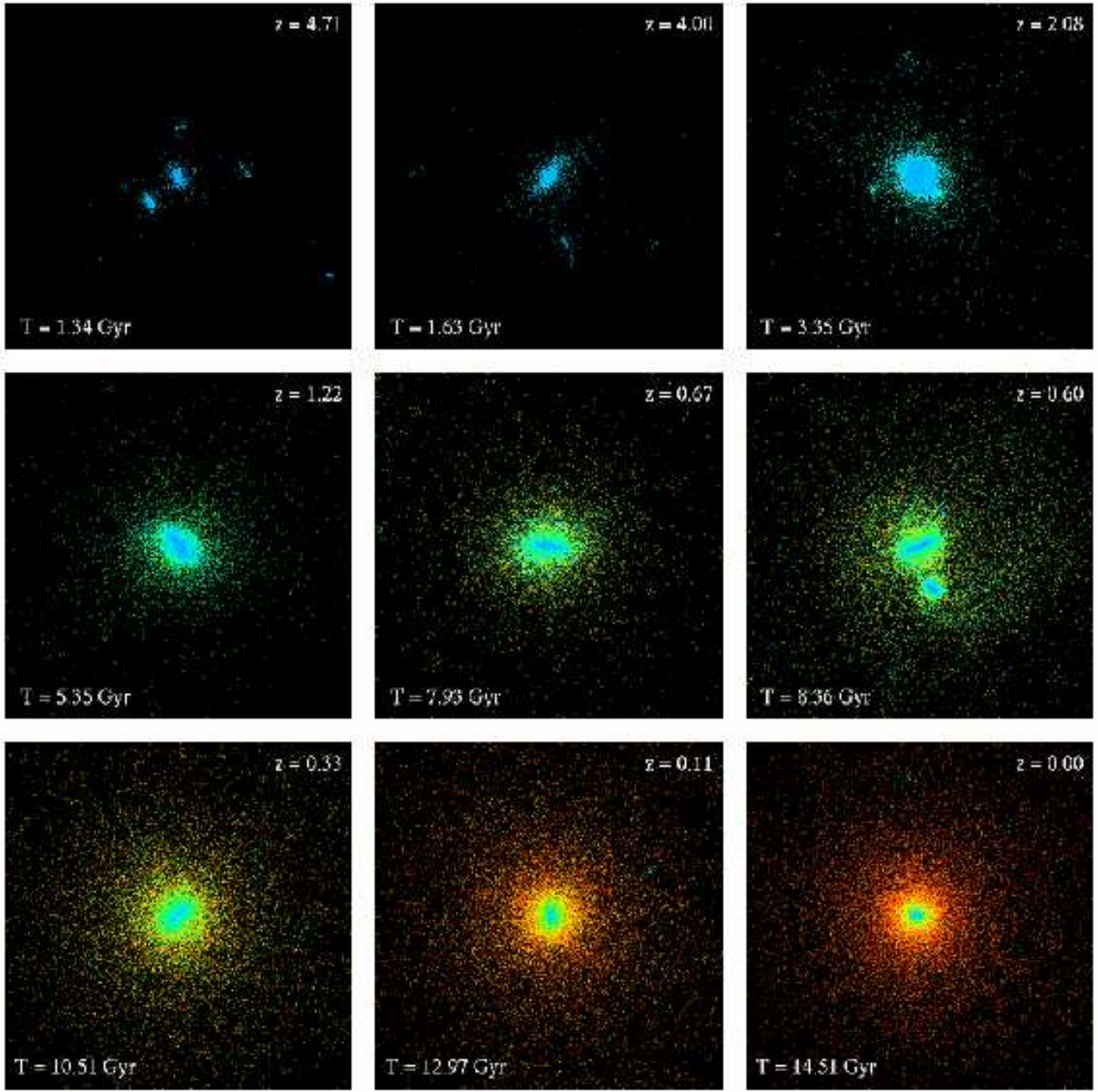}
\caption{Star particles within a cube of $40$ physical kpc on a side,
shown at different redshifts and projected so that the luminous
component of the galaxy at $z=0$ is seen approximately face-on.  Each
particle is colored according to its age at the time shown. Blue and red
correspond to $\tau \lesssim 4$ Gyr and $\tau \gtrsim 10$ Gyr,
respectively. Note that the plotting routine overlays young particles
over old and thus somewhat overemphasizes the importance of young
stellar components. \label{figs:faceon}}
\end{figure*}

The simulation was performed using GRAPESPH, a code that combines the
hardware N-body integrator GRAPE with the smoothed-particle
hydrodynamics (SPH) technique (Steinmetz 1996). GRAPESPH is a fully
three-dimensional, Lagrangian technique optimally suited to study the
formation of highly nonlinear systems in a cosmological context. The
version used here includes the self-gravity of gas, stars, and dark
matter; a three-dimensional treatment of the hydrodynamics of the gas;
Compton and radiative cooling; the effects of a photoionizing UV
background (Navarro \& Steinmetz 1997); as well as a simple recipe for
transforming gas into stars.

The numerical recipe for star formation, feedback and metal enrichment
is similar to that described in Steinmetz \& M\"uller (1994, 1995; see
also Katz 1992; Navarro \& White 1993). In this scheme, star particles
are created in collapsing regions that are locally Jeans unstable, at a
rate controlled by the local cooling and dynamical timescales,
$\dot{\rho}_{\ast}=c_\ast \, \rho_{\rm gas}/\max(\tau_{\rm
cool},\tau_{\rm dyn})$.The parameter $c_\ast$ effectively controls the
depletion timescale of gas in regions of high density, where most star
formation takes place and where $\tau_{\rm dyn}\gg\tau_{\rm cool}$. Star
particles are only affected by gravitational forces, but they devolve
$\sim 10^{49}$ ergs (per solar mass of stars formed) into their
surrounding gas over the $\sim 3 \times 10^7$ yrs following their
formation. This energy input---a crude attempt to mimic the energetic
feedback from evolving massive stars and supernovae---is mainly invested
into raising the internal energy (temperature) of the surrounding gas.
However, because stars form in high-density regions, where cooling
timescales are short, this feedback energy is almost immediately
radiated away.

In order to allow for more effective feedback we assume that a certain
fraction of the available energy, $\epsilon_v$, is invested in modifying
the kinetic energy of the surrounding gas. These gas motions can still
be dissipated through shocks, but on longer timescales, reducing the
star formation efficiency and extending the timescale for the conversion
of gas into stars. Our star formation recipe thus involves two free
parameters, $c_\ast$ and $\epsilon_v$, which we set by matching the star
formation rate of isolated disks to those observed in the local universe
(Kennicutt 1998). We have adopted $c_\ast=0.033$ and $\epsilon_v=0.05$
in the simulation reported here. This is a conservative choice in the
sense that the resulting feedback is relatively inefficient at
regulating star formation: the rate at which gas cools and accretes at
the center of dark matter halos effectively determines the star
formation rate.

\begin{figure*}[tb]
\centering\includegraphics[height=0.73\textheight,clip]{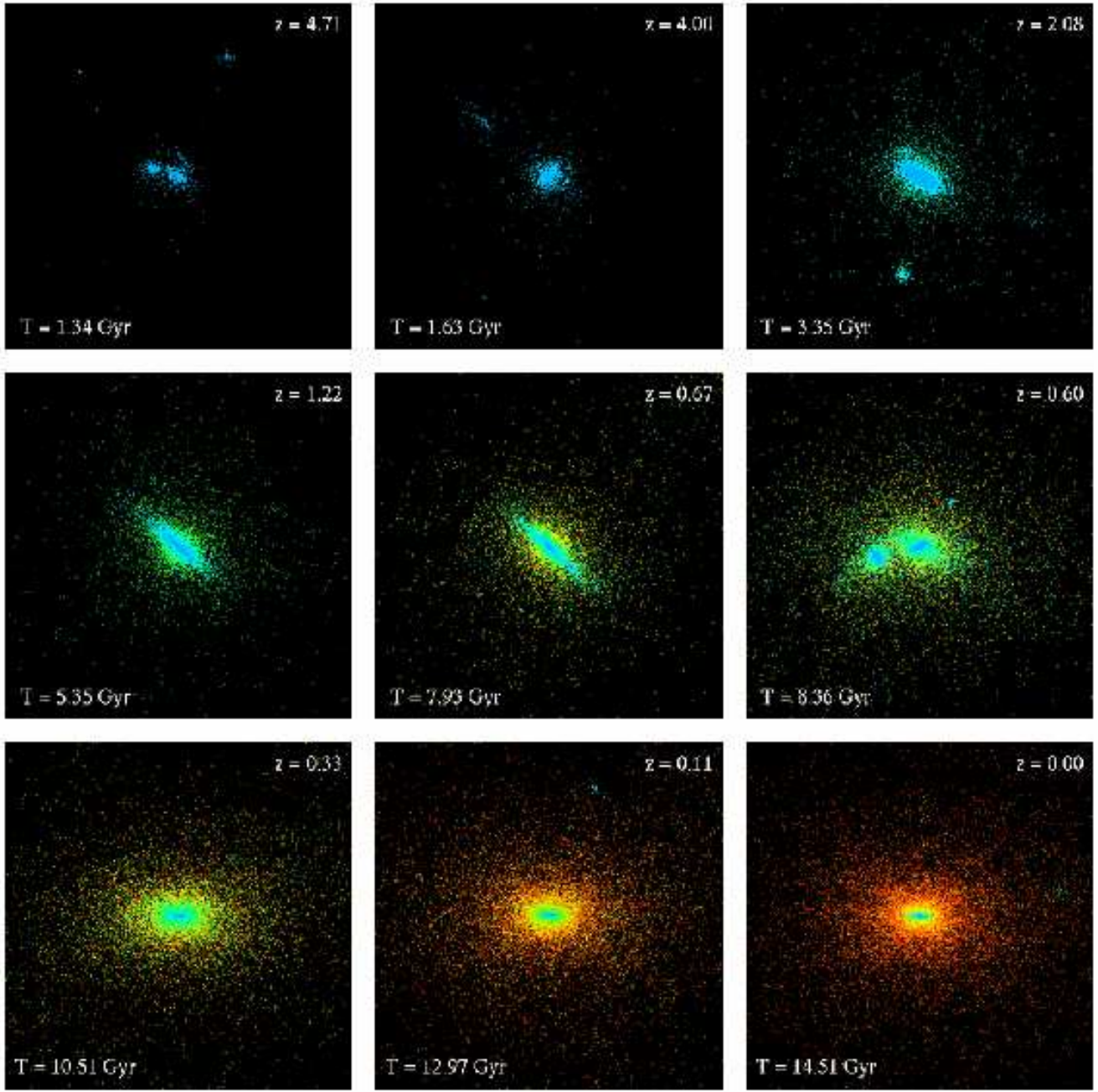}
\caption{As in Figure~\ref{figs:faceon}, but rotated 90 degrees so that the
galaxy at $z=0$ is seen approximately edge-on. \label{figs:edgeon}}
\end{figure*}

\subsection{Initial Conditions}

The simulation assumes the popular ``concordance'' $\Lambda$CDM
scenario; a flat, low-density universe ($\Omega_0 = 0.3$,
$\Omega_{\Lambda}=0.7$) where the mass is dominated by Cold Dark Matter,
with a modest baryonic contribution ($\Omega_b=0.019\,h^{-2}$). The
present day expansion rate is parameterized by
$h=H_0/(100\,\mbox{km\,s$^{-1}$\,Mpc$^{-1}$})=0.65$. The $\Lambda$CDM
power spectrum is normalized so that the linear present-day rms mass
fluctuations on spheres of radius $8\,h^{-1}$ Mpc is $\sigma_8=0.9$.

We simulate a region that evolves to form at $z=0$ a dark halo of mass
$\sim 3 \times 10^{12}\, M_{\odot}$. This halo is selected from a
cosmological simulation of a periodic $32.5\, h^{-1}$ Mpc box and
resimulated at higher resolution, including the full tidal field of the
original calculation. Details on this procedure may be found in Katz \&
White (1993), Navarro \& White (1994) and Navarro \& Steinmetz (1997,
2000).  The mass of gas and dark matter particles is initially
$1.28\times 10^{7}\, M_{\odot}$ and $7.24\times 10^{7}\,$ $M_\odot$,
respectively. The (Plummer) gravitational softening lengthscale is fixed
at 0.5 (physical) kpc.

\begin{figure}[htb]
\centering\includegraphics[width=\linewidth,clip]{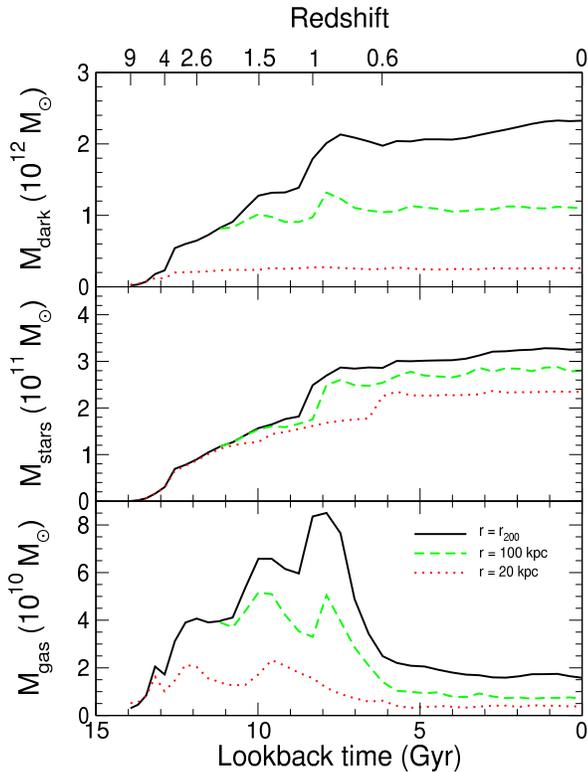}
\caption{The mass evolution of the dark matter, stellar and gas
components of the simulated galaxy, measured within different spherical
shells centered on the main luminous progenitor. Solid lines correspond
to the mass within the virial radius, $r_{200}$, dashed lines to the
mass within 100 (physical) kpc, and dotted lines to the mass within the
luminous radius, $r_{\rm lum}=20$ (physical) kpc. \label{figs:massev}}
\end{figure}

\begin{figure}[htb]
\centering\includegraphics[width=1.3\linewidth,clip]{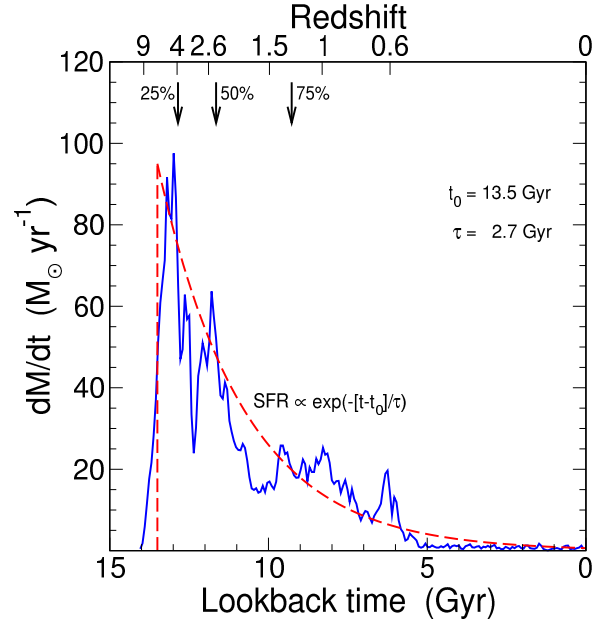}
\caption{Age distribution of all stars within the luminous radius at
$z=0$.  Downward-pointing arrows indicate the time of formation of the
first $25\%$, $50\%$, and $75\%$ of the stars, respectively. Major
mergers are typically associated with short bursts of star formation
that show as peaks in the age distribution.  Overall, the age
distribution is well approximated by an exponential law of age
$t_0=13.5$ Gyr and e-folding timescale $\tau=2.7$ Gyr (see dashed line).
This is comparable to what is usually assumed in spectrophotometric
modeling of the spectral energy distribution of elliptical galaxies
(Guiderdoni \& Rocca-Volmerange 1987; Bruzual \& Charlot 1993).
\label{figs:agehist}}
\end{figure}

\subsection{Analysis}

Galaxy luminosities are computed by simply adding the luminosities of
each star particle, taking into account the time of creation of each
particle and using the latest version of the spectrophotometric models
of Contardo, Steinmetz, \& Fritze-von Alvensleben (1998). We assume that
the stellar Initial Mass Function (IMF) is independent of time.
Quantitative measures quoted in this paper assume a Scalo (1986) IMF
truncated at $0.1$ and $100 \, M_{\odot}$, respectively, unless
explicitly stated otherwise. The spectrophotometric modeling requires a
measure of the metallicity of the stars. GRAPESPH tracks (crudely) the
metal content of each gaseous particle by assuming that each ``star''
devolves a total of $1.7\, M_{\odot}$ of enriched material (per $100\,
M_{\odot}$ of stars formed) to the surrounding gas during the $3\times
10^7$ yrs following its creation.  These metals are dispersed amongst
the gas particles neighboring the star, and their metal content is in
turn inherited by the star particles they spawn. Our treatment of metal
enrichment neglects diffusion processes and is not intended to reproduce
the Fe-rich enrichment by type Ia supernovae, which is thought to occur
on a much longer timescale.  Given its crudeness, we restrict our use of
the metallicity of gas and stars to the spectrophotometric modeling of
the stellar populations. It is our intention, however, to include in
future work a self-consistent treatment of metal enrichment by type I
and II supernovae in order to tap the potential of heavy element
abundance patterns in judging the success of the modeling. This is an
important task, as it has been argued that it might be difficult to
reproduce the alpha element-enhanced abundance of ellipticals in
hierarchical scenarios of formation (see, e.g., Thomas 2001 and
references therein).
 
\begin{figure*}[htb]
\centering\includegraphics[height=0.4\textheight,clip]{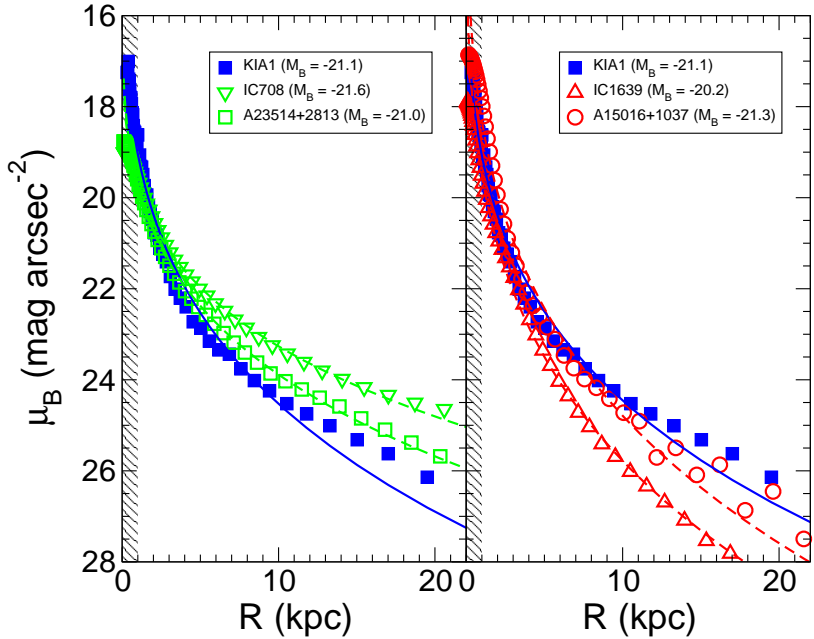}
\caption{B-band surface brightness profile of the simulated galaxy
(labeled KIA1) as a function of projected radius $R$. The left panel
compares it with two ``normal'' elliptical galaxies of similar
luminosity from the Nearby Field Galaxy Survey of Jansen et al. (2000).
The right-hand panel compares KIA1 with two ``compact'' ellipticals from
the same survey. Clearly, the simulated galaxy is much more concentrated
than normal ellipticals of comparable luminosity, and bears a closer
resemblance to compact ellipticals. Lines through the symbols illustrate
the best $R^{1/4}$ fit for each profile. \label{figs:sfbprof}}
\end{figure*}

\section{Results}
\label{sec:results}

\subsection{The Assembly of an Elliptical Galaxy}
\label{ssec:assembly}

As described in detail by Steinmetz \& Navarro (2002), the formation of
the galaxy is characterized by episodes of smooth accretion and by a
number of merger events, some of which are clearly visible in
Figures~\ref{figs:faceon} and~\ref{figs:edgeon}. These figures show the
stellar component within a $40$ (physical) kpc box centered on the most
massive progenitor at different times. Star particles are colored
according to age (blue and red correspond to ages $\lesssim 4$ and
$\gtrsim 10$ Gyr, respectively). Figure~\ref{figs:faceon}
(\ref{figs:edgeon}) shows the system projected along an axis chosen so
that the galaxy at $z\sim 0$ is seen face-on (edge-on).

The morphology of the galaxy varies substantially in response to changes
in the dominant mode of mass accretion. Mergers between progenitors of
comparable mass are frequent at early times. One of these events is
clearly visible in the top left panels of Figures~\ref{figs:faceon}
and~\ref{figs:edgeon}. Major mergers mix stars into a spheroid, around
which smoothly accreting gas may subsequently wrap to form a new
disk-like component.  At $z\sim 2$ the simulated galaxy resembles a
bright Sa/Sb spiral: over $70\%$ of the rest-frame optical luminosity
comes from a roughly exponential disk while the rest comes from a
spheroidal component whose spatial distribution may be well approximated
by an $R^{1/4}$-law.

The disk continues to grow gradually until it is tidally shaken by a
massive satellite (seen in the $z=2.08$ panel of
Figures~\ref{figs:faceon} and~\ref{figs:edgeon}) into a clearly-defined
barred pattern. The bar is resilient and survives the accretion (at
$z=1.6$) of the satellite that triggered its formation.  At $z\sim 0.6$,
the galaxy undergoes a further and final major merger with another disk
galaxy about one third as massive.  During the collision, the remaining
gas ($\lesssim 8\%$ of the total baryonic mass) is efficiently funneled
to the center, where it is quickly converted into stars (Barnes \&
Hernquist 1991). This last episode of star formation is effectively over
by $z=0.5$, leaving only a small ($\sim 1$ kpc diameter) `core' of
younger, metal-rich stars, surrounded by a large spheroid of older
stars. Morphologically, the $z \sim 0.6$ merger turns the galaxy from a
centrifugally supported disk into a slowly-tumbling spheroid that
quickly relaxes into the final configuration which is the subject of the
study we report here.

Quantitatively, the mass assembly process is summarized in
Figure~\ref{figs:massev}, where we plot the evolution of the mass in the
dark matter, gas, and stellar components within various radii. From top
to bottom, the curves in Figure~\ref{figs:massev} correspond to the mass
within the virial radius\footnote{We define the virial radius as the
radius where the mean inner density contrast (relative to the critical
density for closure) is 200; i.e., ${\bar \rho}(r_{200})=200 \times 3\,
H^2(z)/8 \pi G$. We parameterize the present value of Hubble's constant
as $H(z=0)=H_0=100\, h\,$ km s$^{-1}$ Mpc$^{-1}$. All distance-dependent
quantities assume $h=0.65$.}, $r_{200}$; within $100$ (physical) kpc;
and within the ``luminous'' radius, $r_{\rm lum}=20$ (physical) kpc. All
of these curves show that the assembly of the galaxy is essentially
complete by $z=0.6$, just after the last major accretion event. Indeed,
the mass within $r_{\rm lum}$ remains approximately constant after that
time, although that within $r_{200}$ continues to increase, mainly due
to the time-dependent definition of the virial radius. The $z=0.6$
merger leads to a short starburst that depletes (into stars) the last
$\sim 4 \times 10^{9} \, M_{\odot}$ of gas within the merging galaxies.
The final stellar mass of the galaxy is essentially fixed at this event,
since little gaseous material is added to the galaxy afterwards; stars
age passively for the remaining $\sim 7$ Gyr of the evolution. The
galaxy thus has no opportunity to re-form a stellar disk, and its
morphology at $z=0$ is consistent with that of elliptical galaxies. Some
of the properties of the galaxy at $z=0$ are summarized in
Tables~\ref{tab:gxprop} and \ref{tab:gxphot}.

\begin{deluxetable*}{rrrccc}
\tablecaption{Main properties of simulated galaxy at
$z=0$\label{tab:gxprop}} 
\tablewidth{\linewidth}
\tablehead{
\colhead{Radius} &
\colhead{$M_{\rm tot}$} & 
\colhead{$M_{\rm dm}$} & 
\colhead{$M_{\rm stars}$} &
\colhead{$M_{\rm gas}$} &
\colhead{$V_{\rm circ}$} 
\\ 
\colhead{(kpc)} &
\colhead{($10^{11}\, M_{\odot}$)} & 
\colhead{($10^{11}\, M_{\odot}$)} & 
\colhead{($10^{11}\, M_{\odot}$)} &
\colhead{($10^{10}\, M_{\odot}$)} &
\colhead{(km s$^{-1}$)} 
\\ } 
\startdata 
$r_{200}$ & $26.66$ & $23.25$ & $3.26$ & $1.58$ & $195.3$ \\
$200$     & $21.49$ & $18.31$ & $3.06$ & $1.21$ & $214.9$ \\
$100$     & $13.87$ & $11.00$ & $2.80$ & $0.71$ & $244.2$ \\
$ 20$     & $ 4.99$ & $ 2.60$ & $2.35$ & $0.38$ & $327.5$ \\
$2R_{e}$  & $ 2.05$ & $ 0.54$ & $1.51$ & $0.11$ & $578.2$ \\
$R_{e}$   & $ 1.45$ & $ 0.36$ & $1.08$ & $0.06$ & $687.8$ \\
\enddata
\tablecomments{All quantities are measured within the radii given in
column [1] and assume $h=0.65$. The virial radius at $z=0$ is
$r_{200}=300.5$ kpc. The mass per particle is $m_{\rm dm}=7.24\times
10^7 \, M_{\odot}$ for the dark matter. In total, there are $32090$ dark
matter particles within $r_{200}$. Initially (before any star formation
takes place) the gas particle mass is $1.28 \times 10^7\, M_{\odot}$. 
There are $87669$ star particles within $r_{200}$, not all of the same
mass; the average star particle mass is $3.72 \times 10^6 \, M_{\odot}$.
The effective radius at $z=0$ is $R_e=1.32$ kpc, measured in the
$B$-band.}
\end{deluxetable*}

\begin{deluxetable*}{crccccc}
\tablecaption{Photometric properties of simulated galaxy \label{tab:gxphot}}
\tablewidth{\linewidth}
\tablehead{
\colhead{Band} &
\colhead{$L_{\rm tot}$} &
\colhead{$M$} &
\colhead{$R_{e}$} &
\colhead{$\mu^{\rm fit}_{0}$} &
\colhead{$R^{\rm fit}_{e}$} &
\colhead{$\langle\mu\rangle_{e}$} 
\\ 
\colhead{} &
\colhead{[$10^{10}\, L_{\sun}$]} &
\colhead{[mag]} &
\colhead{[kpc]} &
\colhead{[mag arcsec$^{-2}$]} &
\colhead{[kpc]} &
\colhead{[mag arcsec$^{-2}$]} 
\\ 
}
\startdata
$U$ & $ 3.32$ & $-20.7$ & $1.34$ & $12.2$ & $1.78$ & $18.5$ \\
$B$ & $ 4.16$ & $-21.1$ & $1.32$ & $11.8$ & $1.75$ & $18.1$ \\
$R$ & $ 6.65$ & $-22.6$ & $1.29$ & $10.2$ & $1.72$ & $16.5$ \\
$K$ & $19.80$ & $-25.0$ & $1.17$ & $ 7.7$ & $1.58$ & $13.9$
%
%
\tablecomments{Luminosities and radii assume face-on projection,
$h=0.65$ and a Scalo IMF with an upper and lower mass cutoffs of $100$
and $0.1 \, M_{\odot}$, respectively.}
\end{deluxetable*}

The evolutionary features discussed above are easily recognizable in the
age distribution of stars (Figure~\ref{figs:agehist}). Most stars
identified within $r_{\rm lum}$ at $z=0$ are old; the first $25\%$ of
the stars form before $z\sim 4$; the median formation redshift is $z\sim
2.4$ and only $2\%$ of the stars have formed in the past $5$ Gyr. The
dashed line in Figure~\ref{figs:agehist} shows that, overall, the age
distribution compares well with that expected from an exponentially
declining star formation rate, ${\dot M}_{*}\propto e^{-t/2.7{\rm
Gyr}}$, initiated $13.5$ Gyrs ago. These parameters are not dissimilar
from those used to model elliptical galaxies spectrophotometrically
(e.g., Guiderdoni \& Rocca-Volmerange 1987; Bruzual \& Charlot 1993), so
we expect the broad-band colors and other major photometric indicators
of the simulated galaxy to be in reasonable agreement with those of
elliptical galaxies. The numerical resolution achieved in this
simulation (at $z=0$ there are more than $65000$ stars within the
luminous radius, $r_{\rm lum}$) allows us to examine in detail the
structure of the simulated galaxy and to assess whether the
morphological likeness to an elliptical at $z=0$ is further
substantiated by the detailed photometric and kinematic properties of
the galaxy. We turn our attention to this issue next.

\subsection{Photometric Properties}
\label{ssec:photprop}

\subsubsection{Surface brightness profile}
\label{sssec:sfbprof}

\begin{figure}[htb]
\centering\includegraphics[width=\linewidth,clip]{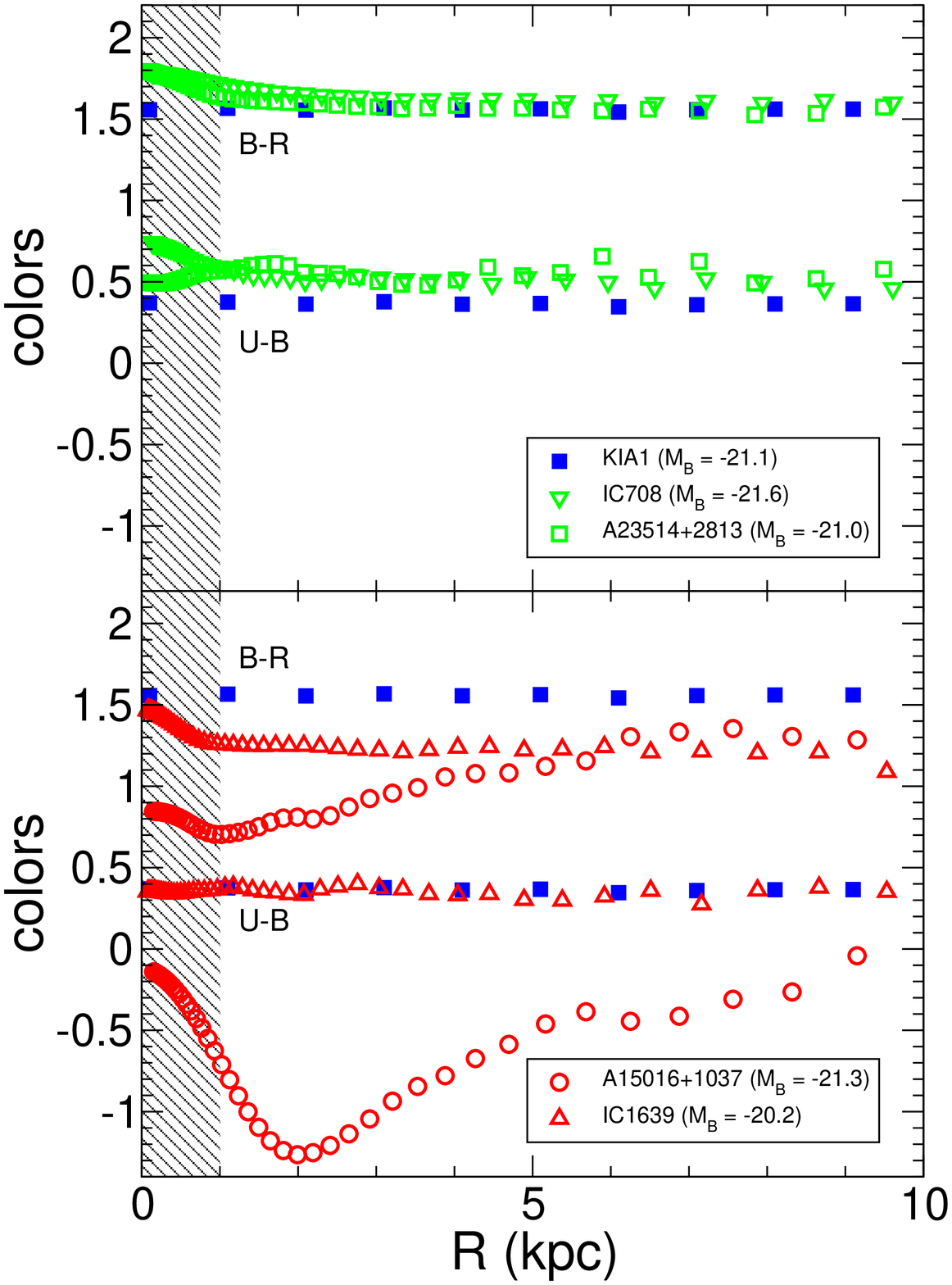}
\caption{$U-B$ and $B-R$ color profiles for the simulated galaxy,
compared with those of the ``normal'' (top panel) and ``compact''
(bottom panel) ellipticals shown in Figure~\ref{figs:sfbprof}.
Observational data is from Jansen et al. (2000). \label{figs:colprof}}
\end{figure}

The face-on surface brightness profile (computed using circularly
symmetric apertures) of the simulated galaxy (labeled KIA1 for short) at
$z=0$ is shown in the left panel of Figure~\ref{figs:sfbprof} (solid
symbols) and compared with two elliptical galaxies of similar $B$-band
absolute magnitude selected from the Nearby Field Galaxy Survey of
Jansen et al. (2000). There are no rescalings allowed in this figure and
therefore the mismatch between simulated and observed ellipticals is
genuine: the simulated galaxy is significantly more concentrated than
IC708 (open triangles) and A23514+2813 (open squares). This is confirmed
by the effective radii of these galaxies: in the $B$-band the effective
radius of KIA1 is $R_e=1.3$ kpc, but those of IC708 and A23514+2813 are
$R_e=11.3$ and $5.8$ kpc, respectively. The simulated galaxy thus
resembles examples of the much rarer class of M32-like ``compact''
ellipticals, such as A15106+1037 or IC1639, shown in the right panel of
Figure~\ref{figs:sfbprof}. These are galaxies of unusually small
effective radii (or unusually high surface brightness) for their
luminosity; the effective radii are $R_e=1.2$ and $1.7$ kpc for
A15106+1037 and IC1639, respectively, comparable to that measured for
KIA1.  

The high concentration of the stellar component may be traced to the
copious star formation that accompanies at high redshift the dissipative
collapse of the galaxy's progenitors. These progenitors are extremely
dense and compact, and their density is inherited by the final remnant.
It is not clear at this time whether this is a generic prediction of
this cosmogony or of a simulation procedure where feedback affects only
minimally the timing and mode of gas cooling and accretion. Indeed,
results reported by other groups (Governato et al. 2002; Sommer-Larsen
et al. 2002a,b), indicate that it is likely that a different
implementation of star formation and feedback--e.g., one that prevents
more effectively the onset of star formation in early collapsing
progenitors--might bring simulations into better agreement between the
observed structure of normal galaxies.

\subsubsection{Colors}
\label{sssec:colors}

The numerous merger episodes that characterize the formation of the
simulated galaxy lead to a fairly uniform stellar component with little
sign of radial gradients in stellar age. This is reflected in the rather
uniform color profiles shown in Figure~\ref{figs:colprof}. This figure
compares the $B-R$ and $U-B$ color profiles of KIA1 with four
ellipticals shown in Figure~\ref{figs:sfbprof}. Because of its old age,
the colors of the simulated galaxy are quite red, comparable to those of
normal ellipticals (top panel in Figure~\ref{figs:colprof}).  On the
other hand, the colors of compact ellipticals are typically bluer and
show greater variety than those of normal ellipticals. Furthermore,
observed ellipticals show a mild propensity for reddish cores which is
not reproduced in the simulated galaxy.  It is likely, however, that
such a tendency is a result of metallicity gradients, which would be
poorly captured by our rudimentary numerical treatment of metal
enrichment.

To summarize, the simulated elliptical is morphologically closer to a
compact elliptical than to a normal one, but it appears as if its
stellar population resembles more that of normal ellipticals. Thus, to
bring the simulated galaxy into closer agreement with observation one
would need to either distend the tightly bound stellar structure of the
simulated galaxy without changing its colors or else allow for recent
episodes of star formation in order to account for the bluer colors of
compact ellipticals. Either alternative would require a major revision
to the way in which star formation and feedback are modeled in our
numerical code. This result is similar to that reported by Abadi et al.
(2003a) for a simulated disk galaxy in the $\Lambda$CDM scenario. Since
Abadi et al. used the same numerical code as in the present study, it
appears as if stellar systems that are too concentrated to be consistent
with observation are a generic feature of our simulation procedure. We
intend to explore systematically the sensitivity of this result to our
choice of star formation and feedback algorithm in future papers of this
series.

\subsubsection{Shape and rotation}
\label{sssec:shaperot}
 
\begin{figure*}[htb]
\centering\includegraphics[height=0.55\textheight,clip]{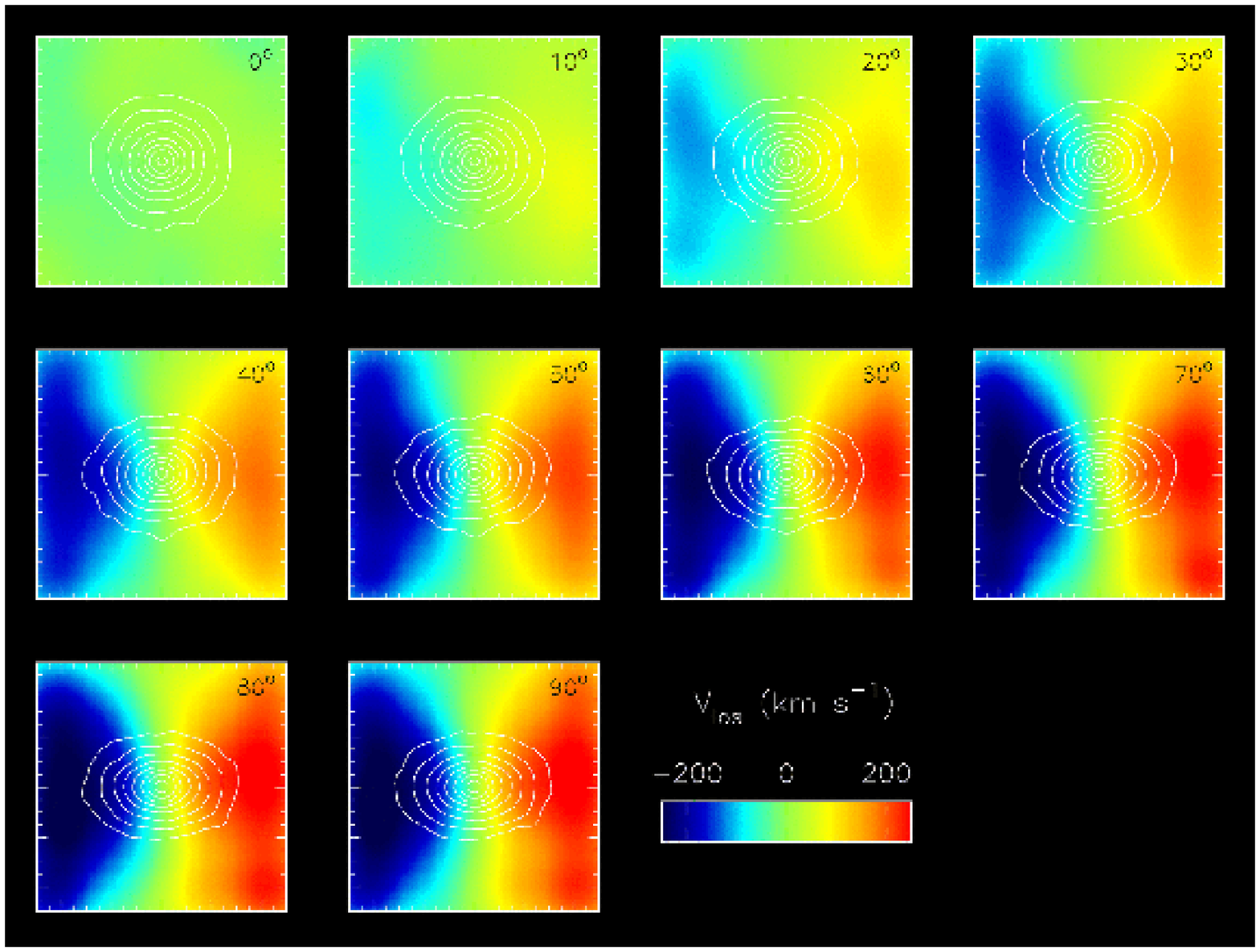}
\caption{Line-of-sight velocity maps and isophotal structure of the
simulated galaxy, measured for ten different projections from $256
\times 256$-pixel images constructed using all star particles in a box
of $10$ kpc on a side. The image is smoothed with a $0.5$-kpc Gaussian
filter matching the gravitational softening length. Ellipses are fit to
these isophotes using the ISOPHOTE routine within the STSDAS package of
IRAF. Orientations are chosen so that the projected major axis is
aligned horizontally in each image. Note the good agreement between the
photometric and kinematic axes, reflecting the very mild triaxial
structure of the simulated galaxy. \label{figs:velmap}}
\end{figure*}

Figure~\ref{figs:velmap} contrasts the photometric structure of the
simulated galaxy with a map of the line-of-sight mean velocity field
along 10 different projections. Each panel corresponds to $10$ kpc boxes
projected so that the inclination of the galaxy increases successively
by $10$ degrees from top-left (face-on projection) to bottom-right
(edge-on view). The orientation is chosen so that the projected major
axis is aligned horizontally in all panels. By construction, the
apparent ellipticity of the galaxy increases along the panel sequence,
from $\epsilon=1-b/a \approx 0.1$ when seen face-on to a maximum
ellipticity of $\epsilon \approx 0.4$ when seen edge-on. The stellar
distribution is thus only mildly triaxial, with axis ratios of order
$1$:$0.9$:$0.6$. This is confirmed by the good alignment between the
projected photometric and kinematic principal axes: the maximum
misalignment is less than $8$ degrees (for the $30\degr$ panel in
Figure~\ref{figs:velmap}).
 
\begin{figure}[htb]
\centering\includegraphics[width=1.35\linewidth,clip]{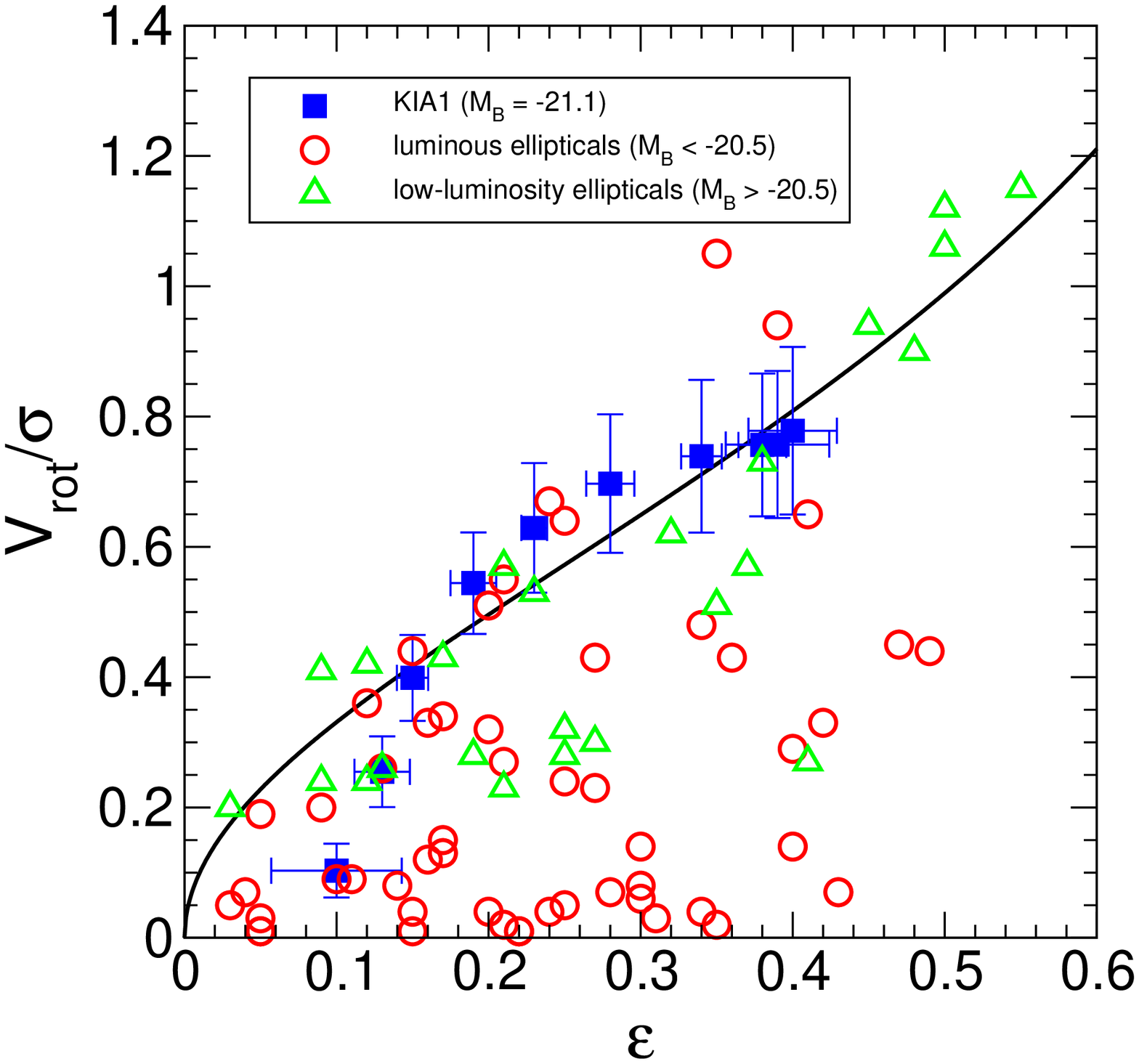}
\caption{The rotation measure, $V_{\rm rot}/\sigma$, plotted versus the
ellipticity of the galaxy. Both quantities are computed at the effective
radius. Solid squares correspond to the same $10$ projections shown in
Figure~\ref{figs:velmap}. Open circles and triangles correspond to
ellipticals brighter and fainter than $M_{B}=-20.5$, respectively.  The
solid line indicates the relation expected for an oblate rotator whose
shape is determined by the degree of rotational support [see Figure 4-6
of Binney \& Tremaine (1987)].
\label{figs:oblate}}
\end{figure}

As is clear from the velocity maps in Figure~\ref{figs:velmap}, the
galaxy has a well defined sense of rotation, and a shape in projection
that is closely linked to the apparent degree of rotational support. The
velocity difference at $R=2\, R_e$ varies from $134$ km s$^{-1}$ for a
$10$ degree inclination to $468$ km s$^{-1}$ when projected edge-on.
This is the kinematic signature expected for an oblate rotator whose
shape is mainly determined by rotation; a result confirmed by
Figure~\ref{figs:oblate}, where we plot the apparent ellipticity of the
galaxy versus the ratio between the apparent rotation speed and the
velocity dispersion, $V_{\rm rot}/\sigma$, measured at $R=R_e$. The
solid squares correspond to the different projections shown in
Figure~\ref{figs:velmap}.  The thick solid line in
Figure~\ref{figs:oblate} corresponds to an ``oblate rotator'' where
rotation dictates the shape of the stellar distribution (see Figure 4-6
of Binney \& Tremaine 1987).  The agreement between the solid line and
the solid squares provides strong evidence for the importance of
rotation in the structure of the simulated galaxy.  The open symbols in
Figure~\ref{figs:oblate} illustrate the well known result that rotation
plays an important role in faint ellipticals (open triangles) and that
rotational support becomes increasingly weak with luminosity.

The importance of rotation in the simulated galaxy is somewhat
unexpected, as it is usually taken to imply that a large fraction of the
stars have formed following a smooth dissipative collapse. In this case,
however, the formation of $95\%$ of the stars predate the final major
($1$:$3$) merger at $z\sim 0.6$ which transforms the two progenitor
disks into a spheroid. Only $3\%$ of the stars form during the merger
event. Simple oblate rotators can apparently form even under
circumstances as inauspicious as a major merger.

The reason why the merger remnant structure is so kinematically
simple---despite its violent assembly history---appears to be linked to
couplings between the internal structure of the progenitor disks and the
orbital parameters of the merger. Indeed, both disks rotate prograde
relative to the orbit of the encounter, and their inclination with
respect to the orbital plane is modest ($18\degr$ in one case, $2\degr$
in the other). This reflects a high degree of coherence between scales
during the torquing process that endowed the progenitors, as well as the
remnant, with net angular momentum.

It is difficult to conclude from a single example whether this
cosmologically-induced coincidence between rotation on small and large
scales is actually responsible for the structural simplicity of
elliptical galaxies. If confirmed, however, it would help to remove one
major objection to the merger model of elliptical galaxy formation: that
dissipationless mergers lead to highly triaxial remnants with kinematic
structure too peculiar to be consistent with observation (Barnes 1992;
Hernquist 1992, 1993). A formation scenario where major mergers play an
important role in the formation of all spheroid-dominated galaxies, and
where the amount of dissipation involved decreases systematically with
luminosity appears thus a viable way of accounting for the gross
morphological and dynamical properties of ellipticals.

\subsubsection{Boxy and disky isophotal shapes}
\label{sssec:boxy}

\begin{figure}[htb]
\centering\includegraphics[width=\linewidth,clip]{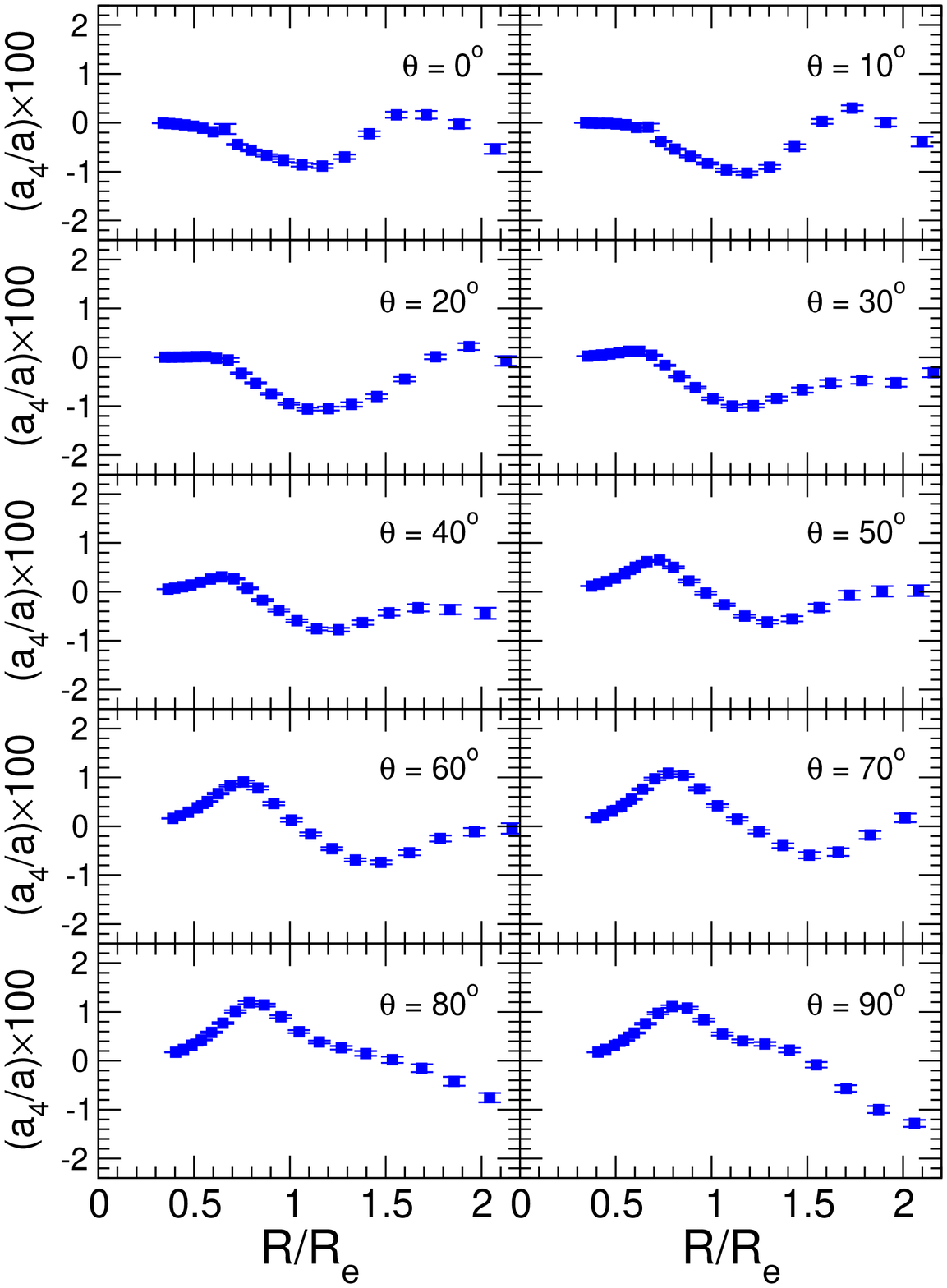}
\caption{The parameter, $a_4$, measuring isophotal deviations from
perfect ellipses, plotted as a function of projected radius for the same
$10$ projections shown in Figure~\ref{figs:velmap}. Low inclination
angles, corresponding to when the galaxy is seen nearly face-on, are
characterized by ``boxy'' ($a_4<0$) isophotes, a trend that reverses for
higher inclinations. Predominantly ``disky'' ($a_4>0$) isophotes are
seen when the galaxy is projected nearly edge-on.
\label{figs:a4prof}}
\end{figure}

\begin{figure}[htb]
\includegraphics[width=1.35\linewidth,clip]{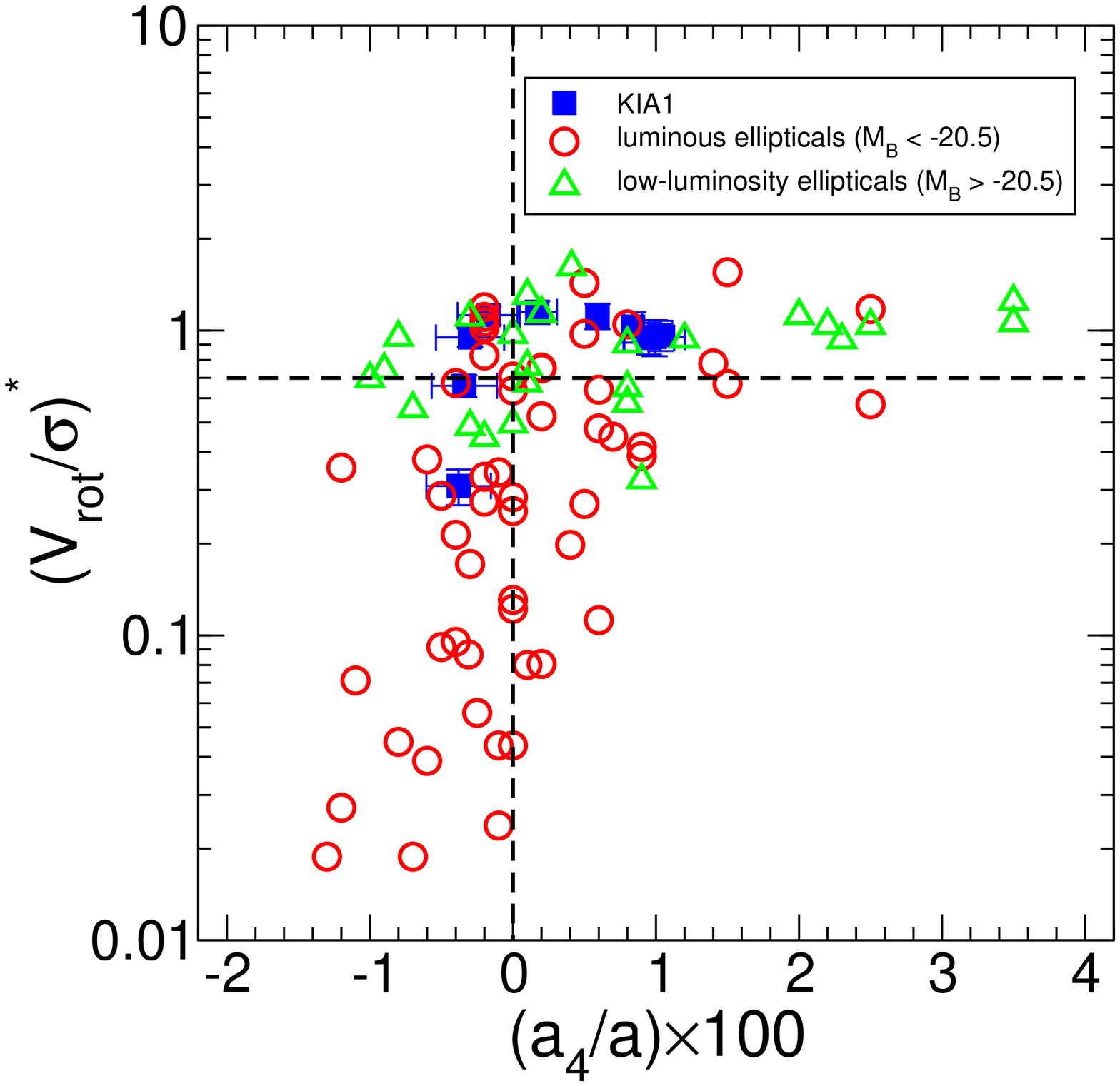}
\caption{Isophotal deviations, as measured by the $a_4$ parameter at the
effective radius, plotted versus the rotational parameter $(V_{\rm
rot}/\sigma)^*$ defined as the ratio between $V_{\rm rot}/\sigma$ and
the value corresponding to an oblate rotator of the same apparent
ellipticity.  Solid squares correspond to the same $10$ projections of
the simulated galaxy shown in Figure~\ref{figs:velmap}. Open symbols
correspond to observed ellipticals from the sample of Bender et al.
(1992). The simulated galaxy seems to follow roughly the trend between
rotation support and diskiness (boxiness) present in the observations.
\label{figs:a4vsig}}
\end{figure}

\begin{figure}[htb]
\centering\includegraphics[width=1.35\linewidth,clip]{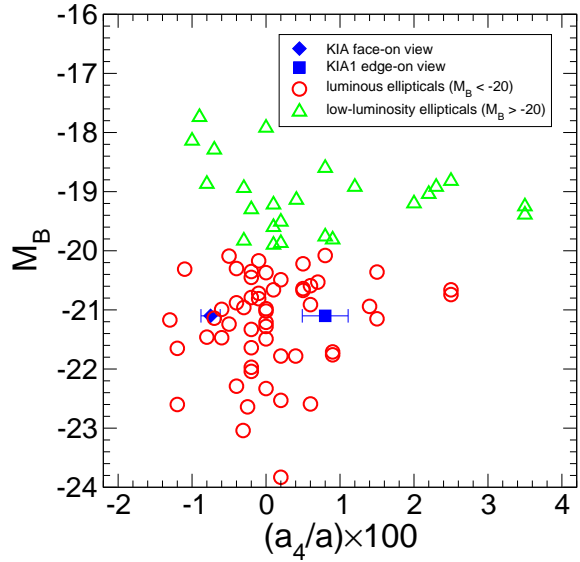}
\caption{Isophotal deviations, as given by the $a_4$ parameter measured
at about the effective radius, plotted versus the absolute magnitude in
the $B$ band. A (weak) trend is seen in observational samples for
fainter galaxies to have diskier isophotes than their brighter
counterparts. The solid symbols illustrate the location of the simulated
galaxy, projected so as to be seen edge-on and face-on. There is no
obvious disagreement between the isophotal deviations measured in KIA1
and in ellipticals of similar luminosity. \label{figs:a4mb}}
\end{figure}

Although the isophotes shown in Figure~\ref{figs:velmap} are well
approximated by ellipses, small but significant deviations from perfect
ellipsoidal shapes are also robustly measured.  Of particular interest
is the $a_4$ parameter, which measures $m=4$ deviations from perfect
ellipses: $a_4<0$ signals ``boxy'' isophotes whereas $a_4>0$ imply
``disky'' deviations from perfect ellipses (Bender \& M\"ollenhoff
1987). Observationally, there is a clear trend between $a_4$ and the
degree to which the galaxy's shape is supported by rotation. Disky
isophotes are typically associated with fast rotators, whereas boxy
ellipticals are most often those exhibiting little rotation (see, e.g.,
open symbols in Figure~\ref{figs:a4vsig}). Given that rotational support
decreases with luminosity (\S~\ref{sssec:shaperot}) it is somewhat
surprising that $a_4$ correlates only weakly with luminosity, as seen in
Figure~\ref{figs:a4mb}. One reason for this is that a number of
fast-rotating ellipticals have boxy isophotes, apparently at odds with
the simple interpretation that disky isophotes are the result of a
dissipative collapse phase responsible for the galaxy's rotational
support.

Our simulation offers a tentative explanation for these trends.
Figure~\ref{figs:a4prof} shows $a_4$ as a function of radius, computed
for the various projections shown in Figure~\ref{figs:velmap}. The
triaxial nature of the stellar component results in $a_4$ values that
vary substantially with inclination and radius. Near the center (i.e.,
for $R\lesssim \, 0.5 R_{e}$), $a_4 \sim 0$, a result which is possibly
associated with smoothing of the image and of the gravitational
potential; a circle of radius $\sim 0.5\, R_{e}$ is of order the
gravitational softening and spans only $34$ pixels. Further from the
center, however, a more clearly discernible inclination-dependent trend
emerges: at the effective radius, isophotes are boxy ($a_4 < 0$) when
the galaxy is seen at low inclinations but the trend reverses sign for
higher inclinations; disky isophotes ($a_4 > 0 $) are seen when the
galaxy is projected nearly edge-on. A trend between $a_4$ and rotation
support thus results, as illustrated by the solid squares in
Figure~\ref{figs:a4vsig}. This figure shows that boxy isophotes may be
obtained even when $V_{\rm rot}/\sigma$ approaches values as high as
those of an oblate rotator; i.e., boxy isophotes may occur even in
systems where rotational support plays an important role. 

We end this discussion with a caveat: the isophotal deviations from
perfect ellipses measured for KIA1 are statistically very significant,
but they are also small, and cover a dynamic range smaller than
observed. In particular, the maximum ``diskiness'' measured in KIA1 is
of order $\sim 1\%$, compared to the $4\%$ reaches in some
low-luminosity systems (Figure~\ref{figs:a4vsig}). Also, the maximum
``boxiness'' in KIA1 does not exceed $a_4 \sim -0.5\%$; whereas $a_4 <
-1\%$ is not unusual in observational samples. As our simulated sample
grows, we expect to be able to elucidate better these trends and to
provide insight into the meaning of the correlations (or lack thereof)
between shape, kinematics, and photometry within the hierarchical merger
model for the formation of ellipticals.

\subsection{Kinematical Properties}
\label{ssec:kinprop}

\subsubsection{Circular velocity profiles}
\label{sssec:vcprof}

\begin{figure}[htb]
\centering\includegraphics[width=1.35\linewidth,clip]{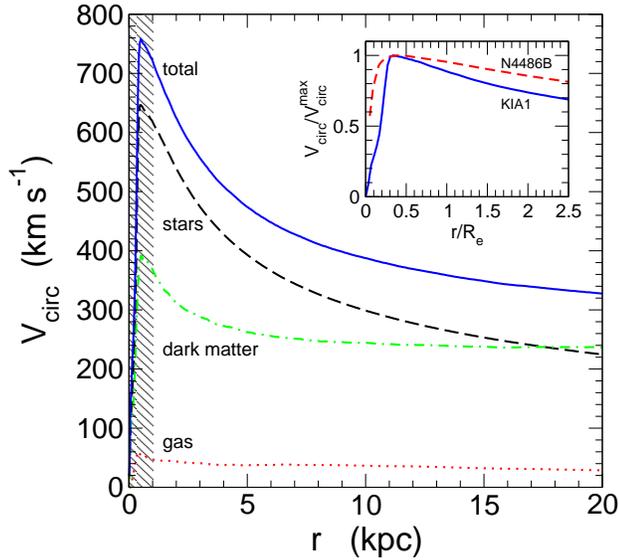}
\caption{Circular velocity profile of the simulated galaxy (solid line),
together with the contributions from the stellar (dashed), dark matter
(dot-dashed), and gas (dotted) components, respectively. The stellar
component dominates over the dark matter as far out as $\sim 17$ kpc
from the center. The circular velocity profile declines steadily from
the center outwards, at odds with the results of Gerhard et al. (2001)
for normal ellipticals, but in closer agreement with the profile of
``compact'' ellipticals. The inset compares the inner circular velocity
profile of KIA1 with the declining profile derived by Gerhard et al. for
the compact elliptical NGC4486B (dashed line). Both profiles have been
scaled to the maximum circular velocity for ease of comparison.
\label{figs:vcprof}}
\end{figure}

The kinematics of the luminous component of the galaxy is largely
dictated by the mass distribution, which may be expressed by the
circular velocity profile, $V_{\rm circ}(r)=(GM(r)/r)^{1/2}$. This is
shown in Figure ~\ref{figs:vcprof}, and indicates that the luminous
(stellar) component dominates strongly the mass distribution near the
center.  Indeed, the stellar component outweighs the dark matter by as
much as $3$ to $1$ at the effective radius and by roughly $1.9$:$1$ as
far away as $5\, R_e$ ($\sim 7$ kpc) from the center. Only beyond $17$
kpc does the dark matter begin to dominate the mass budget in the
galaxy.

The high concentration of the stellar component results in large
circular velocities near the center; at the effective radius the
circular velocity exceeds $700$ km s$^{-1}$. The circular velocity
declines steeply with radius; at $8$ kpc from the center it comes down
to just $\sim 450$ km s$^{-1}$. At face value, this is at odds with the
results of Gerhard et al. (2001), who find that the circular velocity
profile of normal ellipticals---inferred from their line of sight
absorption line profiles---is approximately flat, varying by less than
$10\%$ between $\sim0.2\, R_e$ and $\sim 2 \, R_e$.

Although this discrepancy seems worrisome, one should recall that the
simulated galaxy resembles photometrically a compact elliptical
(\S~\ref{sssec:sfbprof}); a relatively rare class not well represented
in Gerhard et al.'s sample. Indeed, the only compact elliptical in their
dataset is M87's companion NGC4486B, which is referred to by these
authors as ``exceptional in almost all respects'' (see also Prugniel,
Nieto, \& Simien 1987; Peletier 1993). The inset in
Figure~\ref{figs:vcprof} compares the inner circular velocity profile of
KIA1 with that derived by Gerhard et al. (2001) for NGC4486B, each
scaled to the same maximum circular speed and plotted only as far as
$2.5\, R_e$.  The circular velocity profile of NGC4486B is clearly
declining and, over this limited radial range, it does not differ
dramatically from that of KIA1. This is perhaps unsurprising given the
photometric evidence: the central dynamics of such highly concentrated
galaxies is solidly dominated by the luminous component, with a
negligible contribution from the dark matter component. The sharply
concentrated surface brightness profile thus leads to circular
velocities that decline steeply with radius. In less concentrated
ellipticals the dark matter contribution to the mass profile is
proportionally higher, and leads to the ``flat'' circular velocity
curves reported by Gerhard et al. (2001).

\subsubsection{Line-of-sight velocity profiles}

The shape of absorption line profiles contains unique information about
the orbital structure of stars in elliptical galaxies and about the
radial dependence of the gravitational potential. They have been
profitably used in a number of studies to demonstrate the presence of
dark matter within the luminous radius of some elliptical galaxies
(Carollo et al. 1995; Rix et al. 1997; Gerhard et al. 1998; Kronawitter
et al. 2000), as well as to estimate the degree of orbital anisotropy
affecting the global dynamics of spheroid-dominated galaxies.  Much of
this information comes from small but significant departures from
Gaussianity in the velocity distribution of stars along the line of
sight. These deviations are usually quantified by approximating the
line-of-sight velocity distribution (LOSVD) with a Gauss-Hermite series
(van der Marel \& Franx 1993). The dimensionless $h_4$ coefficient of
this series quantifies kurtosis-like deviations, whilst the $h_3$
coefficient provides a dimensionless measure of the skewness of the
distribution.

\begin{figure}[htb]
\centering\includegraphics[width=\linewidth,clip]{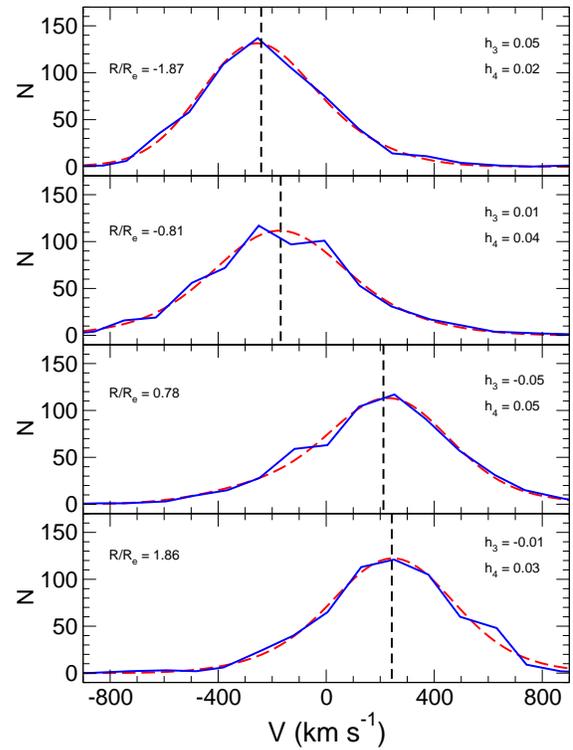}
\caption{Line-of-sight stellar velocity profiles, measured at different
radii along the major axis of an edge-on projection of the simulated
galaxy (solid lines). Dashed lines show the best Gauss-Hermite
polynomial fit, with parameters $h_3$ and $h_4$ as listed in each panel.
Note the anti-correlation in sign between $h_3$ and the mean velocity.
See text for further discussion.
\label{figs:losvd}}
\end{figure}

We have measured LOSVDs in the simulated galaxy by binning particles
along a slit placed on the major axis of different projections of the
simulated galaxy. The bin size is adjusted so as to keep a roughly
constant number of particles within each bin.  Fits using a
Gauss-Hermite series to the LOSVDs measured at four different radii
along the major axis of an edge-on projection of the galaxy are shown in
Figure~\ref{figs:losvd}. As seen in this figure, the velocity
distributions are very nearly Gaussian, although small but significant
deviations are also robustly measured, as quantified by the $h_3$ and
$h_4$ parameters listed in each panel of Figure~\ref{figs:losvd}.

\begin{figure*}[htb]
\centering\includegraphics[width=0.45\linewidth,clip]{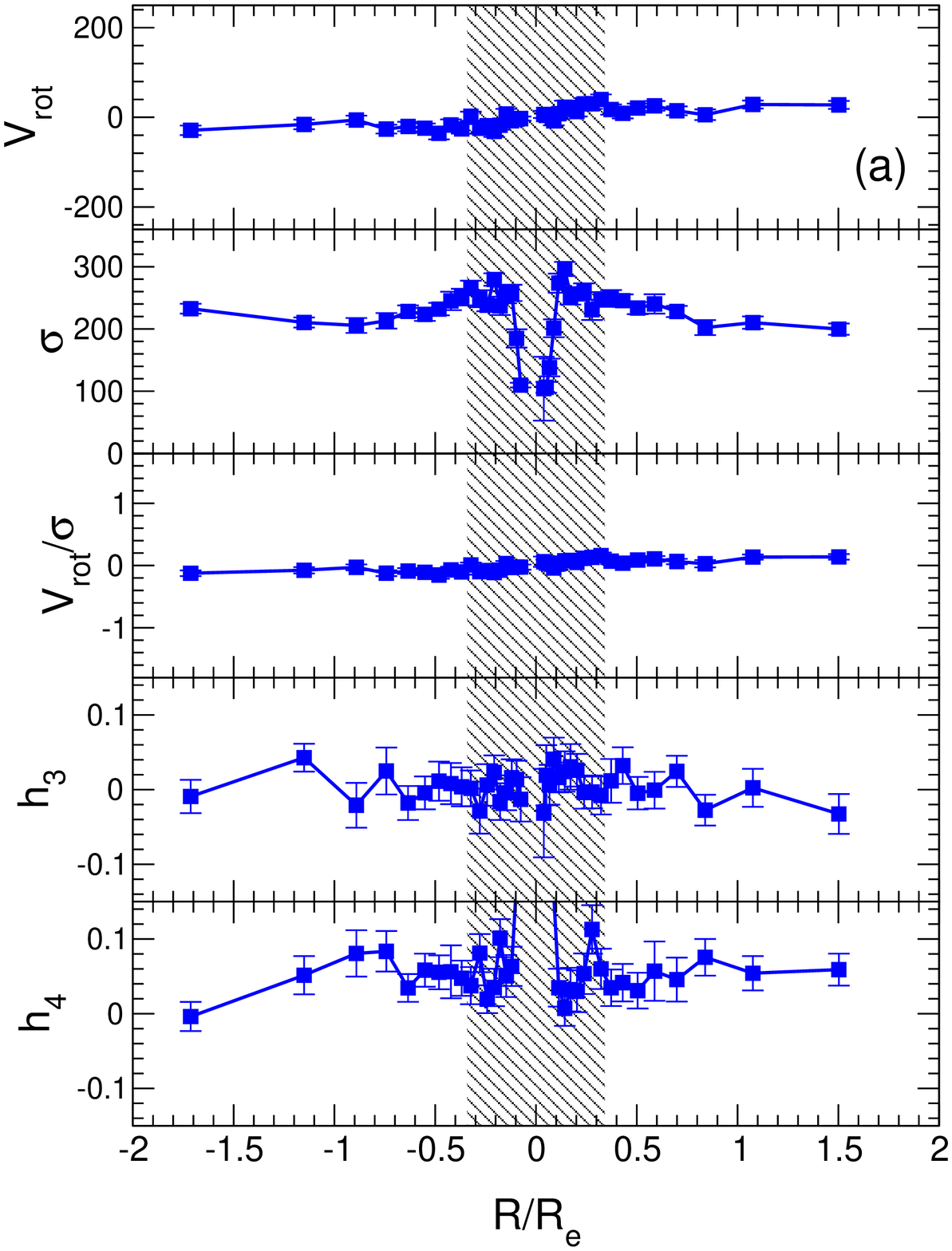}
\hspace{0.3cm}
\centering\includegraphics[width=0.45\linewidth,clip]{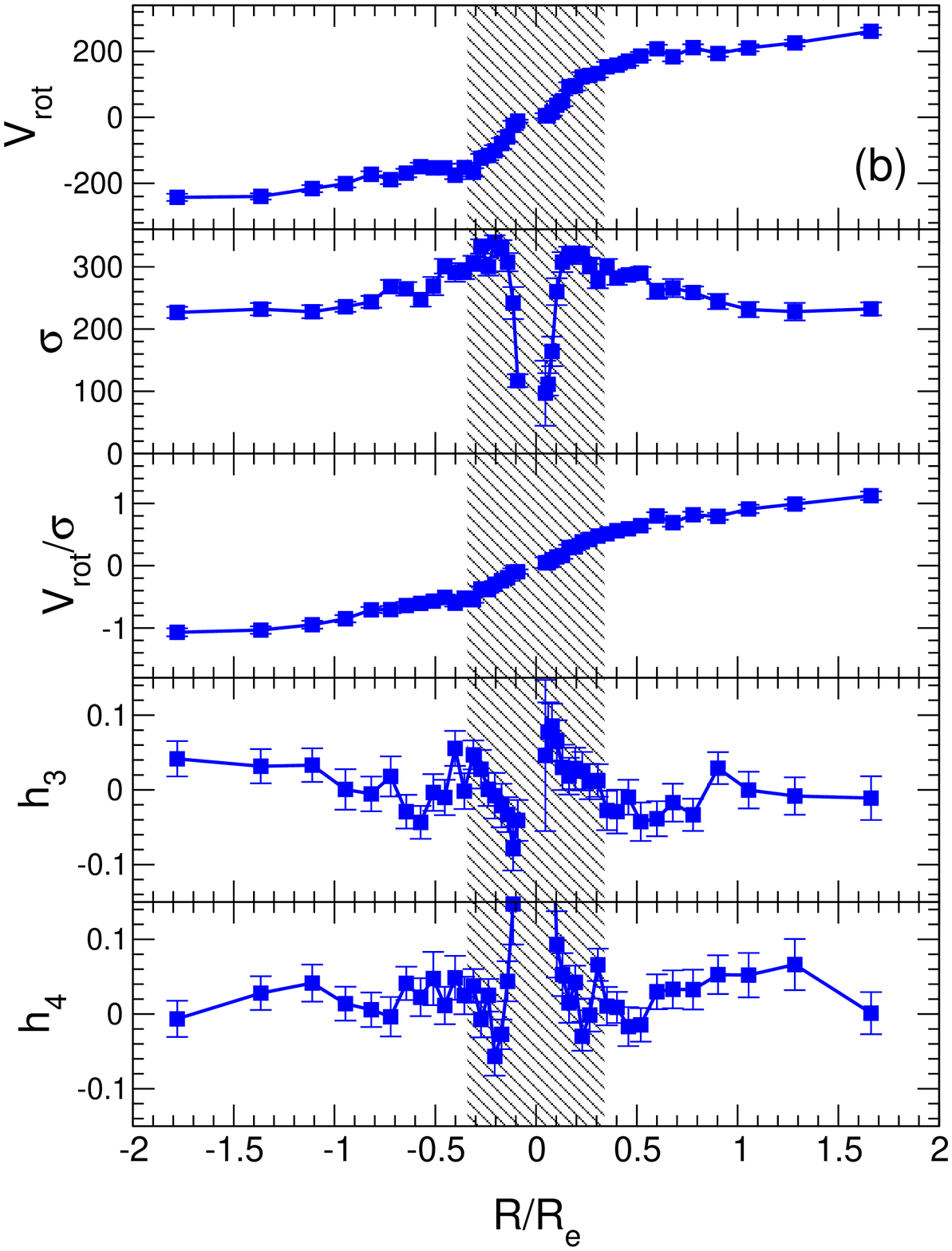}
\caption{Kinematics of the simulated galaxy measured along the major
axis of the (a) face-on and (b) edge-on projections. From top to bottom,
each panel shows: the mean velocity $V_{\rm rot}$; the velocity
dispersion $\sigma$ (both in km s$^{-1}$); the rotational support
measure $V_{\rm rot}/\sigma$; as well as the $h_3$ and $h_4$
Gauss-Hermite parameters. The projected radius $R$ is normalized to the
effective radius $R_e$. Shaded area marks the region likely compromised
by the gravitational softening. \label{figs:ghprof}}
\end{figure*}

The radial dependence of the kinematic structure is synthesized in the
$V_{\rm rot}$, $\sigma$, $V_{\rm rot}/\sigma$, $h_3$, and $h_4$ profiles
shown in Figure~\ref{figs:ghprof} for the case of face-on and edge-on
projections, respectively. This figure highlights a number of
interesting trends. The first one is that the importance of rotation is
a strong function of radius: when seen edge-on, $V_{\rm rot}/\sigma$ is
of order $\sim 0.6$ at the smallest radius confidently resolved in the
simulation ($R\sim 0.5\, R_{e}$) and reaches roughly unity at $R\sim
1.5\, R_{e}$. This is in agreement with earlier work, which found that
the outer regions of merger remnants contain a disproportionate fraction
of the angular momentum of the system (Hernquist 1992, 1993).

\begin{figure}[htb]
\centering\includegraphics[width=\linewidth,clip]{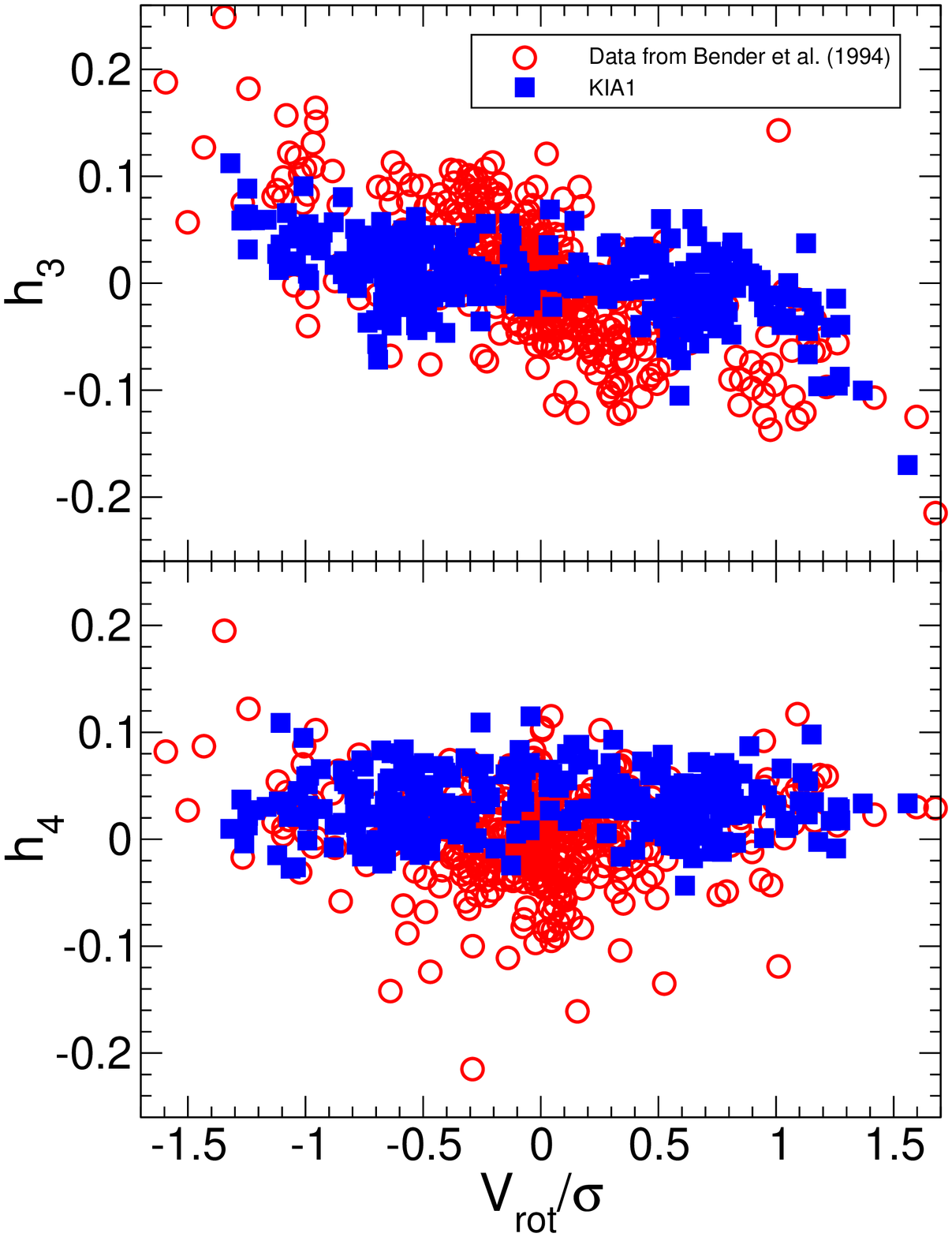}
\caption{Gauss-Hermite parameters $h_3$ and $h_4$ plotted as a function
of $V_{\rm rot}/\sigma$, measured along the major axis of the same 10
projections shown in Figure~\ref{figs:velmap} (solid squares). Open
circles correspond to data from Bender et al. (1994). Note that the
trend between $h_3$ and $V_{\rm rot}/\sigma$ is reproduced in the
simulated galaxy. See text for further discussion.
\label{figs:ghvsigma}}
\end{figure}

Two further interesting trends are apparent in Figure~\ref{figs:ghprof}:
(i) $h_4$ appears to remain positive at essentially all radii, and (ii)
the sign of $h_3$ seems to anti-correlate with that of $V_{\rm
rot}/\sigma$. The first trend implies that LOSVDs are slightly
``boxier'' than Gaussian, a clear signature of the importance of
tangential motions in the outer regions of the galaxy (Gerhard 1993; van
der Marel \& Franx 1993). This is confirmed by the solid squares in the
bottom panel of Figure~\ref{figs:ghvsigma}, where we show the values of
$h_4$ measured on all bins along the major axis of the same $10$
projections shown in Figure~\ref{figs:velmap}: most $h_4$ values are
found to be positive. This trend, however, is not reproduced in the
dataset of Bender et al. (1994) (open circles in
Figure~\ref{figs:ghvsigma}), who find essentially no correlation between
$h_4$ and $V_{\rm rot}/\sigma$, and that $h_4$ in ellipticals is as
likely to be positive as it is to be negative. We interpret this as
reflecting the greater variety of structures prevalent in normal
ellipticals; i.e., not all can be well approximated by simple oblate
rotators. Given that we have a single example, it is not clear how
serious this ``deficiency'' in the model really is; a statistically
significant number of simulations is needed before a more definitive
characterization of this result may emerge.

On a more optimistic note, the correlation between $h_3$ and $V_{\rm
rot}/\sigma$ noted above (and shown in more detail in the top panel of
Figure~\ref{figs:ghvsigma}) reproduces a well-established observational
trend: the asymmetry in the LOSVD is typically such that the excess
``tail'' in the distribution points to the ``origin'' of the velocity
axis (see, e.g., Figure~\ref{figs:losvd}). This asymmetry signals the
importance of coherent rotation in the dynamical structure of the
galaxy; a similar asymmetry would result from a thick rotationally
supported disk and has been used, for example, to argue for massive
gaseous disks as responsible for the velocity structure observed in
damped Ly-$\alpha$ systems (Prochaska \& Wolfe 1998; see, however,
Haehnelt, Steinmetz \& Rauch 1998 for a different interpretation).

As shown in the top panel of Figure~\ref{figs:ghvsigma}, the trend in
the simulated galaxy follows the observed one but is less pronounced
than reported by Bender et al. (1994): at $|V_{\rm rot}/\sigma| \sim 1$
Bender et al. find $|h_3| \sim 0.12$ whereas we find $|h_3| \sim 0.03$
for KIA1. Despite this, it is important to note that the sign of the
observed trend is reproduced in the simulation, in particular since
dissipationless merger simulations have consistently had difficulty
reproducing this trend (Naab \& Burkert 2001).  Our simulation shows
that the mild dissipation that accompanies the merger event at $z=0.6$,
together with the large scale couplings between the internal spin of the
progenitors and the orbital angular momentum of the collision, may be
enough to reconcile such trend with hierarchical merger models of
elliptical galaxy formation.

\subsection{The Fundamental Plane}
\label{ssec:fp}

A powerful diagnostic of the success of simulations in reproducing the
observed properties of galaxies results from comparing their luminosity,
size, and characteristic velocity with the scaling relations linking
these properties in galaxies of various types. The most significant of
these relations for elliptical galaxies is the ``Fundamental Plane''
(Djorgovski \& Davis 1987), or its equivalent, the $D_n$-$\sigma$
relation (Dressler et al. 1987) assiduously used in extragalactic
distance scale studies.

The Fundamental Plane (FP) describes the relation between effective
radius ($R_e$), effective mean surface brightness
($\langle\mu\rangle_e$), and central velocity dispersion ($\sigma_0$) of
early-type galaxies. Clusters and field ellipticals are found to
populate a thin plane in this three-dimensional space, slightly
``tilted'' relative to the plane expected from the virial theorem, as
applied to homologous systems;
\begin{equation}
\log R_e = 2\,(\log\,\sigma_0 +
0.2\,\langle\mu\rangle_e)-\log(\Upsilon/\Upsilon_0) + {\rm constant},
\label{eq:vtre}
\end{equation}
or, equivalently,
\begin{equation}
\log \sigma_0 = \frac{1}{2}\,(\log\,R_e -
0.4\,\langle\mu\rangle_e)+\frac{1}{2}\log(\Upsilon/\Upsilon_0) + {\rm constant}.
\label{eq:vtsg}
\end{equation}

\begin{figure}[htb]
\centering\includegraphics[width=1.35\linewidth,clip]{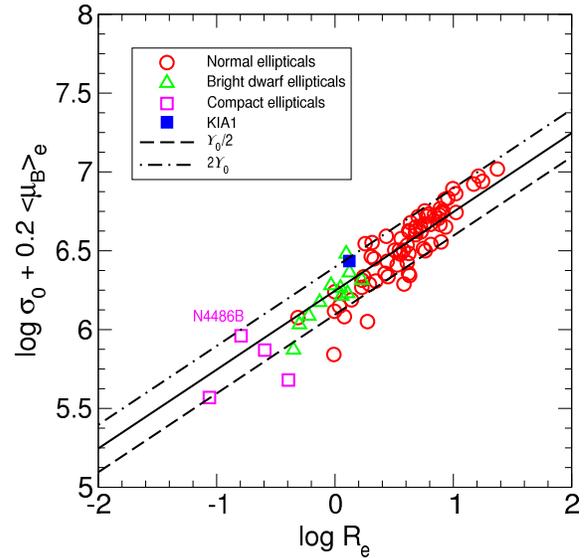}
\caption{Nearly edge-on projection of the Fundamental Plane. Open
symbols are data from Bender et al. (1992). Each symbol corresponds to a
different morphological ``class'' of elliptical galaxy, as labeled in
the figure. Homologous systems obeying the virial theorem and of
constant mass-to-light ratio are expected to populate lines of slope
$1/2$ in this plane, as shown by the solid line. Horizontal deviations
from this line may be ascribed to varying dynamical mass-to-light ratio,
$\Upsilon$. The dashed and dot-dashed lines correspond, respectively, to
systems with $1/2$ and $2$ times the fiducial mass-to-light ratio
parameter $\Upsilon_0$ chosen to draw the solid line. See text for
interpretation. \label{figs:FP1-B-virial}}
\end{figure}

The solid line in Figure~\ref{figs:FP1-B-virial} indicates the loci
expected for systems that follow strictly the homologous virial scaling,
after choosing the reference dynamical mass-to-light ratio,
$\Upsilon_0$, so as to bisect the normal ellipticals in the sample
(shown as open circles). Galaxies to the left of this line have
mass-to-light ratios $\Upsilon > \Upsilon_0$; those to the right have
$\Upsilon < \Upsilon_0$. The tight trend shown in
Figure~\ref{figs:FP1-B-virial} suggests that the Fundamental Plane is
largely a reflection of the virial relations, modulated by relatively
minor variations in mass-to-light ratio.  This is confirmed by the fact
that the best fitting ``plane'' through the dataset shown in
Figure~\ref{figs:FP1-B-virial} is
\begin{equation}
\log R_e = 1.25\,(\log\,\sigma_0 + 0.26\,\langle\mu\rangle_e) -
9.009
\label{eq:vtreobs}
\end{equation}
(Bender et al. 1998). Comparing the quantities between parentheses in
the right-hand side of equations (\ref{eq:vtre}) and (\ref{eq:vtreobs})
confirms that the observed scaling of effective radius with velocity
dispersion and surface brightness is very similar to the virial scaling.

The distribution of normal ellipticals (open circles) around the solid
line in Figure~\ref{figs:FP1-B-virial} illustrates the well-known fact
that the dynamical mass-to-light ratio varies systematically with
luminosity in such systems; large (luminous) ellipticals have higher
dynamical mass-to-light ratios than smaller (fainter) systems. The
spread in mass-to-light ratios is, however, relatively small; most
systems, including morphologically distinct galaxies such as ``bright
dwarf ellipticals'', and ``compact ellipticals'' (following the
nomenclature of Bender et al. 1992), lie between the dashed and
dot-dashed lines in Figure~\ref{figs:FP1-B-virial}, and therefore have
mass-to-light ratios that differ from $\Upsilon_0$ by less than a factor
of $2$.

In order to situate the simulated galaxy (KIA1) within this plane, we
need to estimate its central velocity dispersion, $\sigma_0$, within an
aperture comparable to that used in observations, typically less than
about half the effective radius. Unfortunately, velocities within such
small radii ($\lesssim 0.7$ kpc) in the simulated galaxy are
significantly affected by the gravitational softening; note, for
example, the (artificial) central dip in $\sigma$ shown in
Figure~\ref{figs:ghprof}. We have therefore chosen to estimate
$\sigma_0$ using a de Vaucouleurs model matched to have the same
photometric parameters and to enclose the same total mass within the
effective radius as KIA1. This yields a central velocity dispersion of
$\sigma_0\approx 650$ km s$^{-1}$, a factor of $\sim 2.7$ higher than
would be estimated without accounting for this correction.
 
\begin{figure}[htb]
\centering\includegraphics[width=1.35\linewidth,clip]{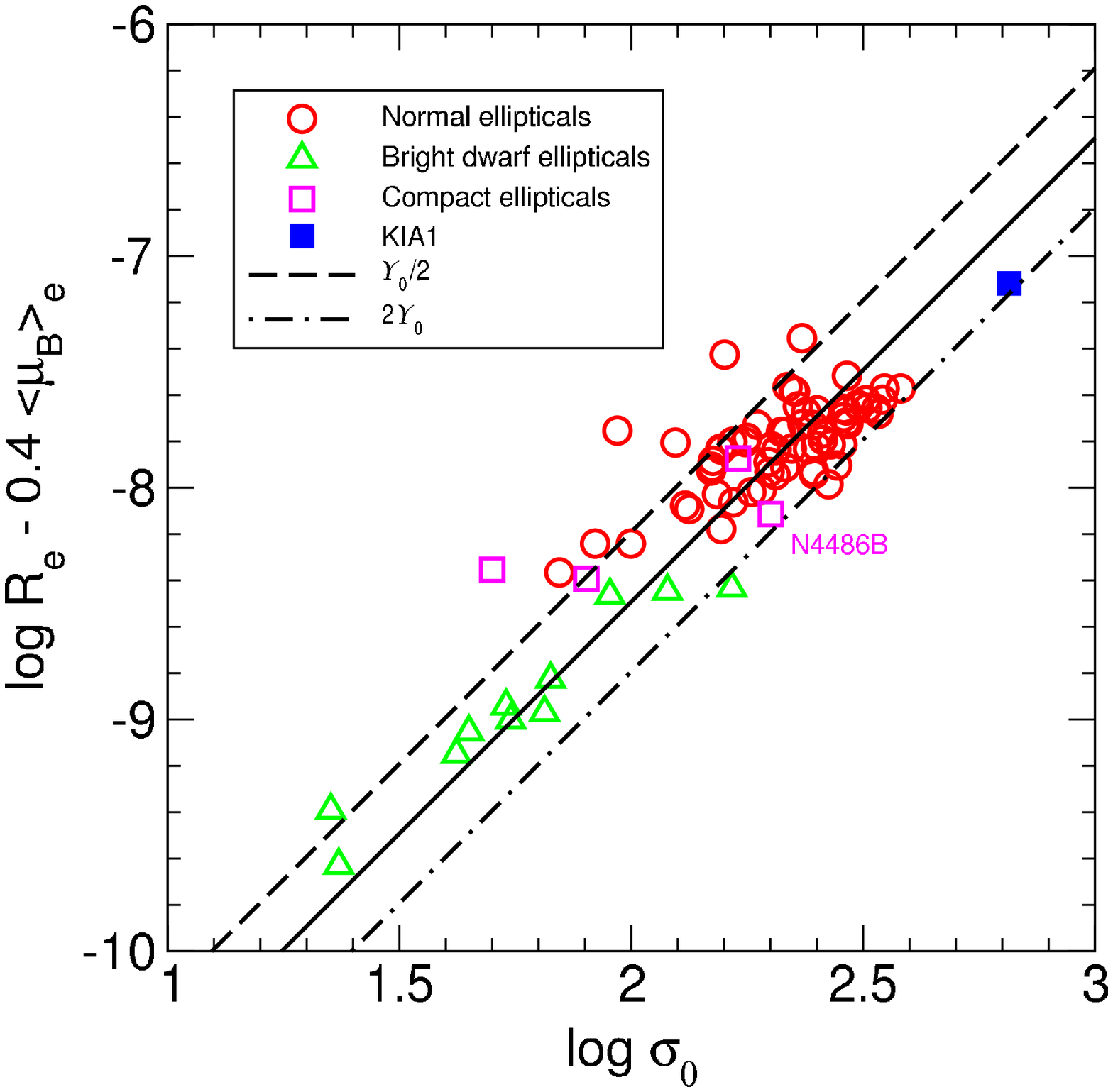}
\caption{Another nearly edge-on projection of the Fundamental Plane. The
axis of the figure are chosen, as in Figure~\ref{figs:FP1-B-virial}, so
that homologous systems of constant mass-to-light ratio obeying the
virial theorem align themselves along lines parallel to the solid line
in the figure, drawn for a fiducial value of the dynamical mass-to-light
ratio, $\Upsilon_0$. Departures to the right or left of this line may be
ascribed to $\Upsilon>\Upsilon_0$ or $\Upsilon<\Upsilon_0$,
respectively. See text for interpretation. \label{figs:FP2-B-virial}}
\end{figure}

Adopting this value of $\sigma_0$, the simulated galaxy sits slightly to
the left of the virial scaling in Figure~\ref{figs:FP1-B-virial},
implying a dynamical mass-to-light ratio of order $\Upsilon \sim 2\,
\Upsilon_0$. This is comparable to that of luminous ellipticals. Indeed,
within one effective radius, $R_e=1.32$ kpc, the total mass is $M_{\rm
tot}(r<R_e)=1.45\times 10^{11} M_{\odot}$, which, combined with a total
luminosity of $L_B=4.16 \times 10^{10} \, L_{\odot}$, results in a total
mass-to-light ratio of order $\Upsilon_B(R_e)\sim 3.5$ in solar units,
comfortably within the range estimated by Gerhard et al. (2001; see
their Figure 13). In other words, the unremarkable mass-to-light ratio
of KIA1 is responsible for its proximity to the Fundamental Plane seen
edge-on.

This result is confirmed in Figure~\ref{figs:FP2-B-virial}, which shows
another nearly edge-on projection of the Fundamental Plane. The lines in
this Figure correspond to the virial scaling expressed in
equation~(\ref{eq:vtsg}), and use the same normalization as those in
Figure~\ref{figs:FP1-B-virial}. Figure~\ref{figs:FP2-B-virial} shows
that KIA1 clearly stands out from the rest: its velocity dispersion,
indeed, is far higher than that of normal ellipticals of similar
luminosity. KIA1 also stands out in terms of its surface brightness, as
is clear from the nearly face-on view of the Fundamental Plane shown in
Figure~\ref{figs:FP3-B-virial}. The surface brightness of KIA1 is
comparable to those of the M32-like group of compact ellipticals, one
example of which is NGC4486B.  

As KIA1, NGC4486B differs from normal ellipticals not only in surface
brightness; it also has a much higher velocity dispersion and circular
velocity than ellipticals of similar luminosity (see, e.g., Kronawitter
et al. 2000 and Gerhard et al.  2001). This is shown in
Figure~\ref{figs:TF}, where we compare KIA1 and NGC4486B with the
``Tully-Fisher relation'' of normal ellipticals derived by Gerhard et
al.  (2001). As for spirals, this relation links the total luminosity of
the galaxy to the (maximum) circular velocity of the system, obtained by
Gerhard et al. through detailed analysis of radially-resolved absorption
line profiles.

Both NGC4486B and KIA1 deviate strongly from the Tully-Fisher relation
of normal ellipticals: they have circular speeds a factor of $3$-$4$
higher than normal ellipticals of comparable luminosity. This seems
consistent with the photometric evidence. Ellipticals as bright as KIA1
have effective radii typically of order $\sim 10$ kpc, a factor of $\sim
10$ times larger than KIA1. Assuming similar stellar mass-to-light
ratios and, as seems likely, that the dark matter plays a minor role
within the effective radius, this implies that KIA1 should have a
circular velocity $\sim \sqrt{10}\approx 3.2$ times higher than average,
in good agreement with the result shown in Figure~\ref{figs:TF}. In
other words, the Tully-Fisher relation for ellipticals derived by
Gerhard et al.  (2001) only applies to ellipticals of ``normal'' surface
brightness, and large deviations from this relation are expected for
systems whose surface brightness deviates strongly from the norm. It
would be interesting to confirm this assertion by extending the same
kind of analysis performed by Gerhard et al. to a sample of compact
ellipticals.

Finally, the comparison shown in Figure~\ref{figs:TF} also suggests that
the higher concentration of the luminous component may be one major
reason why ellipticals of similar luminosity have higher circular speeds
than spirals. The higher the surface brightness the more important the
luminous component in the dynamics of the system and the higher the
circular velocity of the system. In this interpretation, compact
ellipticals offer a rare glimpse at systems where dark matter plays a
negligible dynamical role within the luminous radius. If confirmed, this
property would make them ideal testbeds for models that attempt to
constrain the stellar mass-to-light ratio (and by extension other
properties such as the stellar Initial Mass Function) of
spheroid-dominated galaxies.

\section{Summary}
\label{sec:summary}

We present a detailed analysis of the dynamical and photometric
properties of an elliptical galaxy simulated in the $\Lambda$CDM
scenario. The simulation includes the gravitational and hydrodynamical
effects of dark matter, gas and stars. Star formation is modeled through
a simple recipe that transforms cold, dense, locally-collapsing gas into
stars at a rate controlled by the gas density. Energetic feedback from
stellar evolution is calibrated to match observed star formation rates
in isolated disk galaxy models, but in a form that minimizes the kinetic
coupling between feedback energy and the interstellar medium. As a
result, there is little modulation of the star formation rate, which
tracks closely the rate at which gas cools and condenses at the center
of dark matter halos.
 
Our main results may be summarized as follows:

\begin{enumerate}

\item The galaxy is assembled through a number of high-redshift mergers
followed by brief periods of quiescent accretion which culminate in a
major ($1$:$3$) merger at $z\sim 0.6$. Star formation progresses swiftly
in early collapsing progenitors and accelerates during mergers, only
weakly impeded by our choice of feedback parameters; $50\%$ of the
present-day stars have already formed by $z\sim 2.4$ and $75\%$ of them
by $z\sim 1.3$.

\item The merger at $z=0.6$ mixes the stars of the two disk progenitors
into a highly concentrated spheroidal component that dominates the
galaxy dynamically and photometrically at $z=0$. Few stars form after
this event, and the surface brightness profile of the galaxy at $z=0$
can be reasonably fit with an $R^{1/4}$ law. The effective radius of the
simulated galaxy is $R_e \sim 1.32$ kpc, smaller by a factor of $4$-$8$
than that of normal ellipticals of similar luminosity.  The surface
brightness profile of the simulated galaxy thus resembles that of the
less common class of M32-like ``compact'' ellipticals, although it is
far brighter than most and is not a satellite of brighter galaxies.

\item The luminosity (mass) weighted age of the stars in the galaxy is
$10.0$ ($10.9$) Gyrs, which leads to red colors consistent with normal
ellipticals of similar luminosity, $B-R\sim 1.6$. The late merger that
shapes the spheroid mixes stars very efficiently and results in very
weak color gradients, in agreement with observations.

\item The luminous component dominates the structure of the simulated
galaxy out to $\sim 13$ effective radii, imposing a clearly declining
circular velocity profile: the circular velocity drops from $748$ km
s$^{-1}$ at $r=0.5\,R_e=0.66$ kpc to $578$ km s$^{-1}$ at
$r=2\,R_e=2.64$ kpc. Such sharp decline is at odds with the roughly flat
circular velocity curve of normal ellipticals inferred from detailed
modeling of radially-resolved dynamical observations (Gerhard et al.
2001). However, it appears to be consistent with the mass profile
inferred for compact ellipticals such as NGC4486B.

\item Kinematically, the simulated galaxy resembles an E4 oblate rotator
where rotation support increases sharply with radius: when seen edge-on,
$V_{\rm rot}/\sigma$ increases from $0.6$ at $r=0.5\,R_e=0.66$ kpc to
$1.1$ at $r=1.5\,R_e=1.98$ kpc. The galaxy is only mildly triaxial; its
average axis ratios, as measured from the stellar inertia momentum
tensor within $2\,R_e$ are $b/a \sim 0.9$ and $c/a \sim 0.6$. These
results reflect the importance of dissipation as well as of couplings
between the internal dynamics and orbital parameters of the merger
progenitors. The smooth and kinematically ordered structure of the
simulated galaxy thus lifts (or at least eases) some of the objections
to a merger origin of ellipticals raised by the highly triaxial
structure of the remnants of early dissipationless merger simulations
(Barnes 1992; Hernquist 1992).

\begin{figure}[htb]
\centering\includegraphics[width=1.35\linewidth,clip]{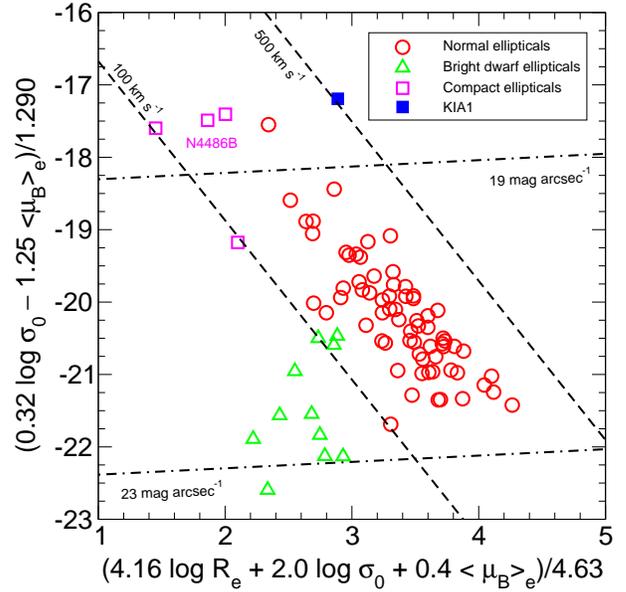}
\caption{Nearly face-on view of the Fundamental Plane, chosen to be
orthogonal to the projections shown in Figures~\ref{figs:FP1-B-virial}
and~\ref{figs:FP2-B-virial}. Surface brightness increases vertically in
this plane, and shows that the simulated galaxy has a much higher
surface brightness than normal ellipticals of comparable luminosity.
\label{figs:FP3-B-virial}}
\end{figure} 

\item Isophotes are well approximated by ellipses, with only a weak
radial variation in position angle and ellipticity. Small but
significant deviations from perfect ellipses are also measured; the
simulated galaxy appears ``boxy'' ($a_4<0$) when seen face-on and
``disky'' ($a_4>0$) when seen edge-on. This appears consistent with the
weak correlation observed between apparent rotation and the sign of the
$a_4$ parameters: boxy ellipticals appear to rotate much more slowly
than their disky counterparts.

\item The line-of-sight velocity distribution of stars (LOSVD) is
approximately Gaussian, although small departures from Gaussianity that
reflect the importance of ordered rotation are also robustly measured.
The Gauss-Hermite coefficient $h_4$ tends to be positive and the sign of
the $h_3$ parameter (anti)correlates with that of the mean velocity.
These trends are reasonably consistent with observation and differ from
those found in earlier dissipationless merger simulations, again
alleviating serious objections to the merger origin of ellipticals
raised by earlier work (see, e.g., Naab \& Burkert 2001 and references
therein).

\item Despite its high concentration, the simulated elliptical lies
close to edge-on projections of the Fundamental Plane. This implies a
dynamical mass-to-light ratio comparable to those of normal ellipticals.
On the other hand, due to its high surface brightness, the circular
velocity (and velocity dispersion) of the simulated galaxy far exceeds
that of normal ellipticals of similar luminosity. It thus clearly stands
out, as other compact ellipticals, from the rest of spheroid dominated
galaxies in face-on projections of the Fundamental Plane or in other
scaling relations such as the Tully-Fisher relation.

\end{enumerate}

\begin{figure}[htb]
\centering\includegraphics[width=1.35\linewidth,clip]{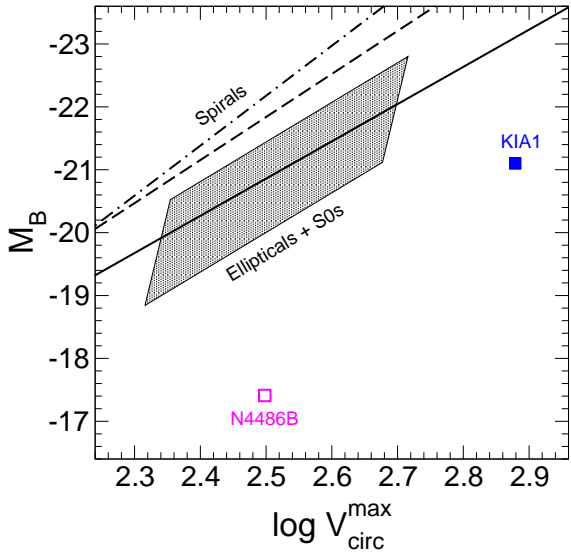}
\caption{The ``Tully-Fisher'' relation of elliptical galaxies (i.e., the
maximum circular velocity plotted versus $B$-band absolute magnitude),
as derived by Gerhard et al. (2001) for a sample of $21$ nearly round
ellipticals (shaded box), compared with the simulated galaxy (KIA1) and
the compact elliptical NGC4486B.  Both galaxies deviate strongly from
the relation that holds for normal ellipticals. This is consistent with
the fact that both NGC4486B and KIA1 are about one order of magnitude
smaller than normal ellipticals but have similar stellar mass-to-light
ratios. \label{figs:TF}}
\end{figure}
Overall, the simulation described here shows that a hierarchical merging
process that includes dissipation and star formation is a promising way
of accounting for many of the observed structural properties of
spheroid-dominated galaxies. Repeated episodes of dissipational collapse
followed by merger events lead to spheroids that are only mildly
triaxial and of relatively simple kinematic structure. This is in better
agreement with observation than earlier merger models between stellar
disks that neglected dissipation and star formation and eases a number
of concerns regarding the viability of merger models for the origin of
ellipticals. We see this as a positive step towards reconciling the
observed structure of elliptical galaxies with a hierarchical assembly
process where mergers play a substantial role.

At the same time, agreement with observation is not perfect: despite its
late assembly redshift ($z\sim 0.6$), the stellar component is more
centrally concentrated than normal ellipticals. This is intriguing, as
it demonstrates that high stellar density does not necessarily imply
high redshift of assembly, as is generally assumed in the ``monolithic
collapse'' model (Partridge \& Peebles 1967; Peebles 2002). However, we
do not consider this to be a robust generic prediction for galaxies
forming in $\Lambda$CDM halos such as the one under consideration here.
Rather, as discussed in earlier papers of this series (Abadi et al.
2003a,b), it reflects the ineffectiveness of our feedback algorithm at
curtailing the efficient cooling of gas and its rapid transformation
into stars in dense, early-collapsing progenitors.

Thus, although several compact ellipticals that closely resemble
photometrically the simulated galaxy may be found in the Nearby Field
Galaxy Survey of Jansen et al. (2000; e.g., IC1639 and A15016+1037), we
cannot claim to have identified the mechanism responsible for the
phenomenon of compact ellipticals. Still, our preliminary conclusion
that such bright compact ellipticals are just ``scaled up'' versions of
M32 (or NGC4486B) requires that they should have extremely high central
velocity dispersions, $\sigma_0\sim 600$ km s$^{-1}$. Observational
estimates of $\sigma_0$ for these galaxies are as yet unavailable, but
they may provide a straightforward way to validate or to rule out some
of our conclusions. This is important, since many of the best known
examples of compact ellipticals are low-luminosity satellites to bright
galaxies, a result that has led to speculation that tides may play an
important role in their formation (Faber 1973; Bekki et al. 2001).
Should IC1639 and A15016+1037 be confirmed to be in the same class as
M32 or NGC4486B would serve to undermine the idea that tidal stripping
plays a crucial role in the formation of these systems (see also Nieto
\& Prugniel 1987; Choi et al. 2002, for other lines of evidence).

It seems clear from this discussion that shedding light on this issue
will require further investigation of other star formation/feedback
implementations, especially since galaxies with unrealistically high
stellar concentration appear to be a general result of the numerical
implementation discussed here. Reversing this trend would require a
feedback algorithm able to hinder star formation in high-redshift
progenitors much more efficiently than in the modeling adopted here.
There is indeed preliminary indication that this may actually be
possible (Springel 2000; Sommer-Larsen et al. 2002a,b; Croft et al.
2002). At the same time, it is also clear that the physically compelling
description of feedback apparently needed is still somewhat beyond
reach. Accounting fully for the structure and dynamics of elliptical
galaxies in a $\Lambda$CDM cosmogony may well have to wait until a
proper description of the interaction between energetic feedback
processes and the interstellar medium finally emerges.

\acknowledgements

This work has been supported by grants from the U.S. National
Aeronautics and Space Administration (NAG 5-10827), the David and Lucile
Packard Foundation, and the Natural Sciences and Engineering Research
Council of Canada.

\end{document}